\documentclass[twocolumn,prx,aps,superscriptaddress,longbibliography]{revtex4-2}

\usepackage{amsmath,amssymb}
\usepackage[english]{babel}
\usepackage[T1]{fontenc}
\usepackage{inputenc}
\usepackage{graphicx}
\usepackage{xcolor}
\usepackage{gensymb}

\usepackage[colorlinks]{hyperref}
\hypersetup{
  linkcolor  = blue,
  citecolor  = blue,
  urlcolor   = blue,
  colorlinks = true,}

\usepackage[commentmarkup=todo, authormarkup=none, markup=default, highlightmarkup=background, todonotes={textwidth=1.5cm, textsize=tiny}]{changes}

\definechangesauthor[name=GK, color=olive]{GK}
\definechangesauthor[name=AP, color=red]{AP}
\definechangesauthor[name=GS, color=green]{GS}

\newcommand{\pheliqs}{Univ. Grenoble Alpes, CEA, Grenoble INP, IRIG, PHELIQS, F-38000 Grenoble, France}

\newcommand{\lncmiG}{Univ. Grenoble Alpes, CNRS, Univ. Toulouse, INSA-T, LNCMI-EMFL, UPR3228, Grenoble, France}
\newcommand{\imr}{Institute for Materials Research, Tohoku University, Oarai, Ibaraki, 311-1313, Japan}

\begin{document}

\title{\textbf{Successive Electronic Topological Transitions in the Antiferromagnet UPd$_2$Al$_3$}}
\author{A.~Gourgout}\email[Present address: Laboratoire des Solides Irradiés, CEA/DRF/IRAMIS, CNRS, Ecole Polytechnique, Institut Polytechnique de Paris, F-91128 Palaiseau, France]{}
\affiliation{\pheliqs}
\author{G.~Bastien}\email[Present address: Lab.~Natl.~Metrol.~\& Essais LNE, 29 Ave Roger Hennequin, F-78197 Trappes, France]{}
\affiliation{\pheliqs}
\author{D.~Aoki }
\affiliation{\imr}
\affiliation{\pheliqs}
\author{G.~Seyfarth}
\affiliation{\lncmiG}
\author{I.~Sheikin}
\affiliation{\lncmiG}
\author{A.A.~Varlamov}
\affiliation{CNR-SPIN, via del Fosso del Cavaliere, 100, 00133 Roma, Italy}
\author{G.~Zwicknagl}
\address{Institut f\"ur Mathematische Physik TU Braunschweig, Mendelssohnstr.~3, 38106 Braunschweig, Germany}
\author{J.~Flouquet}
\affiliation{\pheliqs}
\author{G.~Knebel}
\affiliation{\pheliqs}
\author{A.~Pourret}
\email[E-mail me at: ]{alexandre.pourret@cea.fr}
\affiliation{\pheliqs}

\date{\today }

\begin{abstract}

We report successive anomalies at low temperature in the magnetic field dependence of the thermoelectric signal in the  heavy fermion compound UPd$_{2}$Al$_{3}$ inside the antiferromagnetic state up to the metamagnetic transition at $H_\text{M} =18$~T. Based on renormalisation perturbation theory and the partitioning of the $f$ orbitals into localized and delocalized parts, our analysis attributes these anomalies to complex topological changes of the Fermi surface driven by Zeeman effect.  The observation of a sudden change of sign both in the thermoelectric power and in the Hall coefficient at $H_\text{M}$ in addition to the appearance of large quantum oscillations in the thermoelectric power above $H_\text{M}$ indicate a strong Fermi surface reconstruction at the metamagnetic transition due to the unfolding of the electronic bands. 
\end{abstract}


\maketitle
\section{Introduction}
Quantum phase transitions are characterized by changes in the nature of the ground state, as between magnetic (anti-ferromagnetic (AF) or ferromagnetic (FM)) and paramagnetic (PM) phases, in the zero-temperature limit. The conventional framework for studying such quantum phase transitions focuses on different symmetries and the identification of local order parameters that spontaneously break them at the transition. However, the discovery of distinct phases of matter with the same symmetry has recently introduced a more general notion of order associated with the global topology of ground-state wave functions, as in topological materials (semimetals, insulators and superconductors). This is the case for Lifshitz transitions, which are examples of quantum phase transitions that occur without any change in symmetry \cite{Blanter, Lifshitz}. A Lifshitz transition is characterized by a topological reconstruction of the Fermi surface, such as when two Fermi surface pieces merge to form a single one. In the density of states (DOS), such a transition is characterized by van Hove singularities, arising from the critical points where the Fermi velocity $\frac{d\epsilon(k)}{dk}$ vanishes at the Fermi surface. Such electronic topological transitions, neglected in mean-field theories, have recently been suggested as a driving force to modify the ground-state properties in correlated electron systems such as cuprates \cite{Norman2010,Leboeuf2011}, pnictides \cite{Liu2010, Wang2015} and heavy fermions \cite{Yelland2011b, Pfau2013, Pourret2013_2, Boukahil2014}. Due to the strong quasiparticle renormalization, the latter possess flat bands close to the Fermi level, often accompanied by van Hove singularities in the renormalized DOS. These flat bands are extremely sensitive to external parameters such as doping, pressure or magnetic field. In particular, field induced Lifshitz transitions have been identified in an increasing number of systems. Prominent examples are the PM compound UTe$_2$ \cite{Niu2020}, the AF compound YbRh$_2$Si$_2$ \cite{Pfau2013, Pourret2013_2} and the FM compounds UCoGe \cite{Bastien2016, Leenen2024} and YbNi$_4$P$_2$ \cite{Pfau2017}, where a cascade of field-induced Lifshitz transitions has been observed.

Another kind of field-induced quantum transition is the metamagnetic transition \cite{Kramers}, which corresponds to a sudden and strong increase of the magnetization as a function of an external magnetic field, often through a first-order transition or via a steep crossover. Such metamagnetic transitions are observed both in localized AF systems with strong magnetic anisotropy — typically associated with spin-flop transitions \cite{Stryjewski1977} — and in PM itinerant electron systems that transform into a FM state \cite{Levitin1988}. Metamagnetic transitions generally occur in the presence of strong magnetic fluctuations, which can give rise to the emergence of new quantum phases, as observed, for example, in the $d$-electron metallic system Sr$_3$Ru$_2$O$_7$ with the appearance of a nematic phase \cite{Perry2001, Borzi2007, Lester2015}. The magnetic character of these systems at low field can differ substantially, reflecting the complexity of establishing a unified understanding. The change of the magnetic structure at the metamagnetic transition gives rise to a Fermi surface reconstruction due to the folding (unfolding) of the bands and the change of the magnetic Brillouin zone. For example in the AF compound CeRh$_2$Si$_2$, which shows spin-flop transitions from an AF ordered state to a polarized paramagnetic (PPM) state with a partial localization of the 4$f$ electrons, a strong reconstruction of the Fermi surface is observed at the metamagnetic transition \cite{Settai1997, Abe1997, Knafo2010, Palacio2015}. In CeRu$_2$Si$_2$ a pseudo-metamagnetic transition occurs in the PM state at 7.8~T to a polarized state. This transition is accompanied by both, a change in the magnetic fluctuations from dominant AF to FM ones above the transition and a Fermi surface reconstruction \cite{Haen1987, Flouquet2002, HAoki2014, Boukahil2014}. Recently in the PM superconductor UTe$_2$, the striking change of the carrier density at the metamagnetic transition at $H_M=34.5$~T is interpreted as a dramatic change of the Fermi surface \cite{Niu2020b}. The question whether the Fermi surface reconstruction associated to the collapse or entering of the magnetic order is also a well defined Lifshitz transition remains an open question.

 Here we will focus on the AF heavy-fermion superconductor UPd$_2$Al$_3$, which crystallizes in the hexagonal PrNi$_2$Al$_3$ structure (space group P6/$mmm$), shown in the upper left inset of Fig.~\ref{Fig1} (lattice parameters $a= 5.37$~\AA~ and $c$ = 4.191~\AA). It orders antiferromagnetically at $T_N = 14.5$~K, with moments ferromagnetically ordered in the basal plane and antiferromagnetically along the $c$ axis with a wave vector $\mathbf{Q}$ = (0, 0, 1/2) \cite{Geibel1991, Krimmel1992} (see upper left inset in Fig.~\ref{Fig1}). The sub-lattice magnetization is about $M_0$ = 0.85~$\mu_B$/U \cite{Krimmel1992} indicating that UPd$_2$Al$_3$ is far from any magnetic instability. Superconductivity occurs below $T_{SC} = 2$~K. The Sommerfeld coefficient, extrapolated from the temperature dependence of the specific heat above $T_{SC}$, yields $\gamma = 140$~mJmol$^{-1}$K$^{-2}$ attesting that UPd$_2$Al$_3$ is a moderate heavy fermion system \cite{Geibel1991}. The large specific heat jump at $T_{SC}$ shows that heavy quasiparticles are involved in superconductivity. When magnetic field is applied in the basal plane of the hexagonal structure, i.e. along [10$\bar{1}$0] or [11$\bar{2}$0] (see upper left inset of Fig.~\ref{Fig1}), a first-order metamagnetic transition occurs at $H_M=17.8$~T or $H_M=18.4$~T, respectively \cite{DeVisser1993,DeVisser1994, Oda1994, Sakon2002}. While the Fermi surface in the AF phase has been extensively studied both experimentally \cite{Inada1994, Haga1999, Inada1999, Haga2000, Blackburn2006,Fujimori2007} and theoretically \cite{Sandratskii1994,Knopfle1996,Zwicknagl2003}, the Fermi surface in the PPM state regime above $H_M$ has been investigated by quantum oscillation measurements only in a very narrow field range close to $H_M$ and a strong Fermi-surface reconstruction occurring at $H_M$  has been reported \cite{Terashima1997}. 

 In the present study, we performed thermoelectric power (TEP) measurements with the magnetic field applied within the basal plane (along the [10$\bar{1}$0] direction) up to 29~T. At low temperatures and for fields below $H_M=18$~T, a succession of anomalies in the Seebeck coefficient ($S$) is observed over a broad magnetic field range (4-16~T). Using renormalisation perturbation theory and the partitioning of the $f$ orbitals into localized and delocalized parts \cite{Zwicknagl2003, Zwicknagl2016}, these different anomalies in the AF state have been attributed to complex topological changes occurring in the Fermi surface. In addition, we have observed a sudden change of sign both, in $S$ and in the Hall coefficient, at $H_M$ accompanied by the appearance of large quantum oscillations in the TEP above $H_M$. Most likely, this is the consequence of a topological Fermi surface reconstruction due to the unfolding of the electronic bands and a weakening of the renormalized masses at the metamagnetic transition. Finally, we investigated the angular dependence of the de Haas-van Alphen (dHvA) signal within the AF phase by rotating the magnetic field up to 15~T from [10$\bar{1}$0] to [11$\bar{2}$0]. This allowed us to reveal a new branch of the Fermi surface inside the AF phase. 
 
 The succession of anomalies in the Seebeck coefficient observed here over a broad magnetic field range is reminiscent of the multiple Lifshitz transitions reported in other uranium-based heavy fermion superconductors, such as UCoGe \cite{Bastien2016} and URu$_2$Si$_2$ \cite{palacio2013b}, suggesting that cascades of field-induced changes of the Fermi surface topology - rather than one single abrupt reconstruction - may be a common signature of correlated 5$f$-electron systems approaching a metamagnetic or field-tuned instability. This provides a route to understand how unconventional superconductivity, non-Fermi-liquid behavior, and quantum criticality emerge in the vicinity of field-tuned instabilities in correlated electron systems.

\section{Experimental}

High-quality single crystals of UPd$_2$Al$_3$ were grown by the Czochralski method in a tetra-arc furnace. The single crystal ingot was oriented by taking X-ray Laue photograph and cut using a spark cutter. The quality of the crystal studied here was checked by resistivity measurements down to low temperature using a home-made Adiabatic Demagnetization Refrigerator (ADR). The residual resistivity ratio (RRR=$\frac{\rho (300~K)}{\rho (T_{SC})}$) of the sample was higher than 100. TEP measurements down to 400~mK and up to 29~T were carried out at the high magnetic field facility LNCMI-Grenoble using a \textsuperscript{3}He cryostat with a cold finger. A constant current has been used to create the thermal gradient during the field sweeps. The TEP signal was measured using a standard "1-heater 2-thermometers" setup in vacuum. Thermal gradient was applied along the [11$\bar{2}$0] direction and magnetic field along [10$\bar{1}$0] (see upper left inset of Fig.~\ref{Fig1}). Measurements up to 16 T have been performed using a dilution refrigerator and a superconducting magnet. In contrast to the high magnetic field experiment, here it was possible to increase the magnetic field step by step and averaging at each field to cross-check the absolute values of the TEP with those from the continuous field sweeps. Magnetoresistance and Hall effect have been measured up to 35~T at 50~mK using the top-loading dilution refrigerator of LNCMI Grenoble. In addition, dHvA measurements have been performed in a top-loading dilution refrigerator down to 25~mK and up to 15~T using the field modulation technique in a superconducting magnet with magnetic field rotating from the $c$~axis [0001] to the basal plane ([11$\bar{2}$0] direction), and from the [11$\bar{2}$0] direction towards the [10$\bar{1}$0] direction within the basal plane.

\section{Fermi surface reconstruction at $H_M$}

\begin{figure}[t]
	\begin{center}
		\includegraphics[width=8cm]{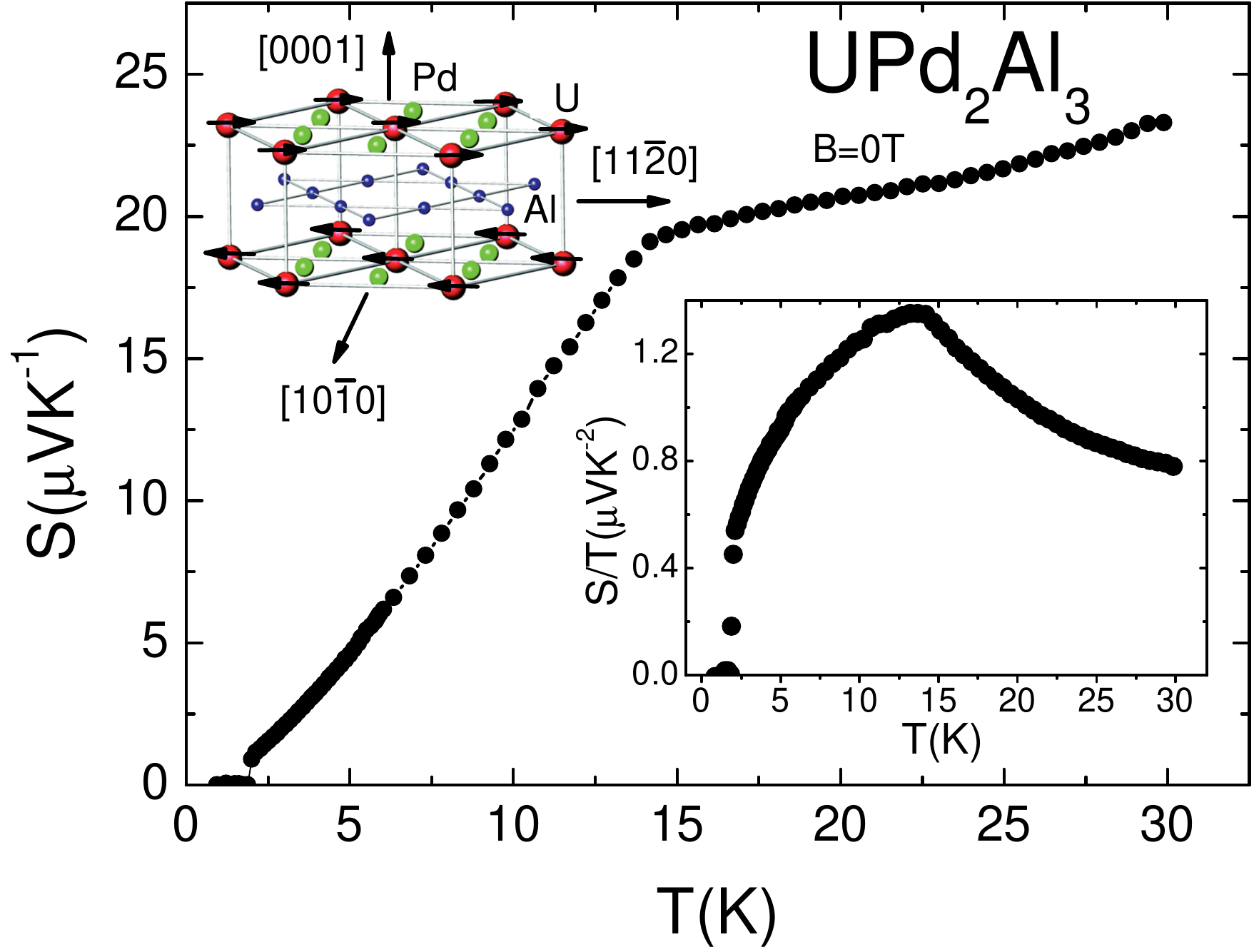}
		\caption{\label{Fig1} (Color online). Temperature dependence of the Seebeck coefficient $S(T)$ at zero magnetic field. The TEP shows a kink at the PM-AF transition, $T_{N}=14.5$~K and a sharp step when entering in the superconducting state below $T_{SC}=2$~K. Upper left inset: Hexagonal crystal structure of UPd$_2$Al$_3$ showing the three main crystallographic orientations discussed within this work. The magnetic moments are ordered ferromagnetically within the (basal) plane orthogonal to the $c$ axis ([0001] direction) and stacked antiferromagnetically along the [0001] direction with an ordering vector $\mathbf{Q}$ = (0, 0, 1/2). Lower right inset: $S(T)/T$ at zero field which never reaches a constant value in the AF state. }
	\end{center}
\end{figure}

Figure~\ref{Fig1} shows the temperature dependence of the TEP, $S(T)$, in UPd$_2$Al$_3$ at zero magnetic field for temperatures below 30~K. A clear kink appears at the Néel temperature, $T_N = 14.5$~K, marking the onset of the AF order, while $S$ drops to zero in the superconducting state below $T_{SC} = 2$~K. The application of a magnetic field produces only minor changes in the temperature dependence of $S(T)$. When plotting $S(T)/T$ in the normal state (see lower right inset of Fig.~\ref{Fig1}), it is evident that $S/T$ never approaches a constant value. This indicates that a Fermi-liquid regime is not realized in the AF state and that magnetic scattering remains dominant.

\begin{figure}[t]
	\begin{center}
		\includegraphics[width=8cm]{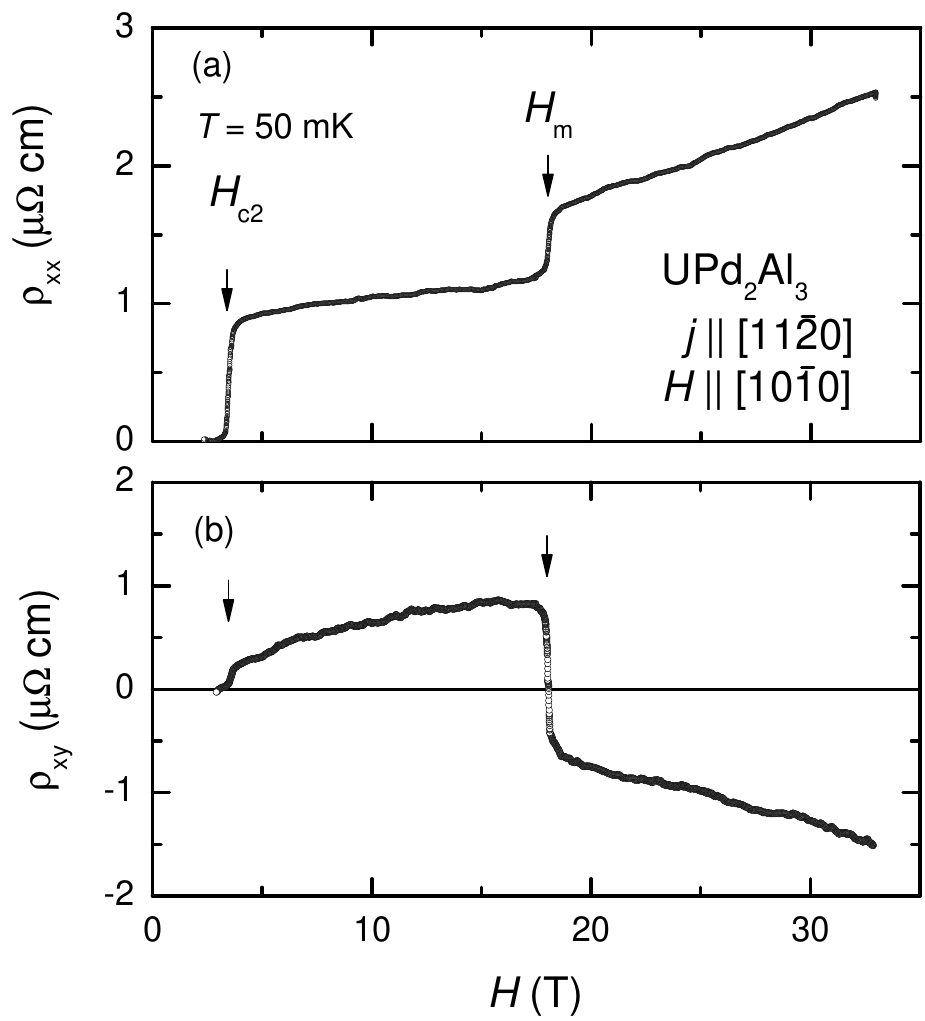}
		\caption{\label{Fig2} (a) Field dependence of the (longitudinal) magnetoresistance, $\rho_{xx}(H)$, and (b) of the (transverse) Hall resistance, $\rho_{xy}(H)$ at $T=50$~mK, exhibiting a strong anomaly and a change of sign at the metamagnetic transition at $H_M$, respectively.}
	\end{center}
\end{figure}  

In Fig.~\ref{Fig2}, we plot the field dependence of the magneto-resistance $\rho_{xx}$ and the Hall resistance $\rho_{xy}$ at 50~mK up to 35~T, with the magnetic field applied along the [10$\bar{1}$0] direction for a current along the [11$\bar{2}$0] direction, i.e. by an angle of $30\degree$ with respect to the magnetic field. The superconducting upper critical field $H_{c2} = 3.5$~T is defined by the midpoint of the transition. Above the superconducting transition, $\rho_{xx}(H)$ increases monotonously up to $H_M = 18$~T.  At $H_M$, $\rho_{xx}$ increases suddenly by 0.4 $\mu \Omega$cm (35~\%), which is the signature of the metamagnetic transition \cite{DeVisser1993,DeVisser1994,Sakon2002}. In previous experiments \cite{DeVisser1994} with current along the [0001] direction a negative jump occurred at $H_M$. Such an anisotropy of the magnetoresistance through $H_M$ occurs also in other systems \cite{Matsuda2000} and is related to spin-dependent scattering, magnetic anisotropy in the material, or a Fermi-surface change \cite{Freericks1992, Knebel2024}. The Hall resistance $\rho_{xy}(H)$ shows an abrupt change of sign at $H_M$ from positive to negative suggesting that the dominant charge carriers change from hole-like in the AF state to electron-like in the PPM state above $H_M$ [see Fig.~\ref{Fig2}(b)]. These prominent features are a first  experimental hint of a possible Fermi-surface reconstruction at $H_M$, involving a change of the dominant charge carriers and also related to the band unfolding when passing from the AF to the PPM state. Further compelling evidence for such a scenario will be discussed later on when combining those results with data from another very sensitive probe which is TEP. 

\subsection{Fermi surface in the AF phase below $H_M$}

\begin{figure}[h!]
	\begin{center}
		\includegraphics[width=9cm]{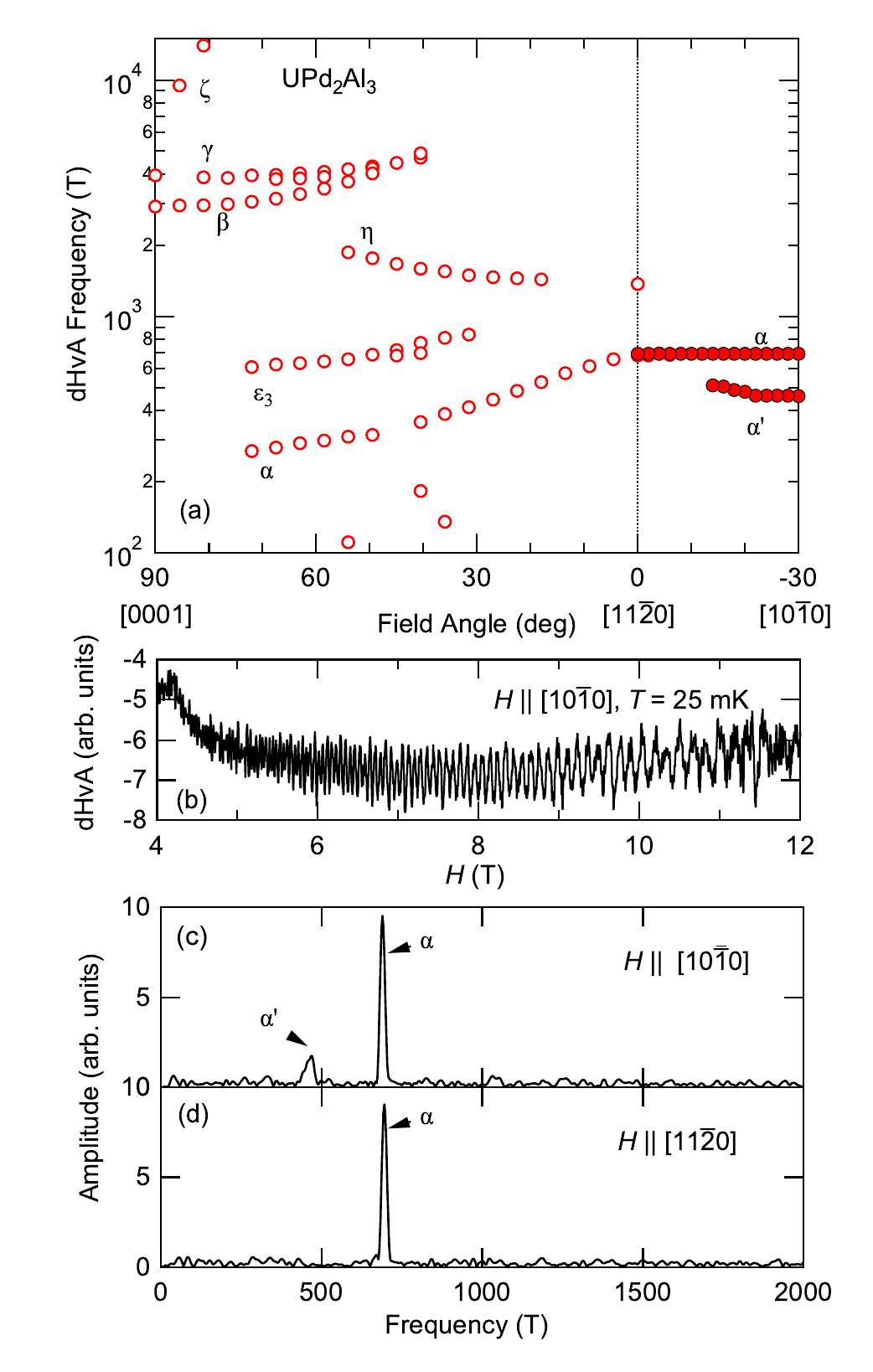}
 \caption{\label{Fig3_ok} (Color online). (a) Angular dependence of the dHvA frequencies, in good agreement with the literature~\cite{Inada1999}, except for the $\alpha'$ branch between the [10$\bar{1}$0] to [11$\bar{2}$0] direction, which has not been observed previously. (b) Field dependence of the dHvA signal from 4 to 12~T (below $H_M$) for H~$\parallel$~[10$\bar{1}$0] at $T=25$~mK. Quantum oscillations are observable within the entire field range. (c) Corresponding FFT spectrum (for $H \parallel$~[10$\bar{1}$0]),  and also for $H \parallel$~[11$\bar{2}$0] (d) .  
 }
	\end{center}
\end{figure}

 The Fermi surface of UPd$_2$Al$_3$ in the AF state had already been studied in detail by Inada et al. \cite{Inada1999}. The angular dependence of the dHvA frequencies in the present study is summarized in Fig.~\ref{Fig3_ok}(a), and in very good agreement with previous data \cite{Inada1999}, in particular concerning the rotation from the $c$~axis to the [11$\bar{2}$0] direction. However, our detailed measurements within the basal plane from $[11\bar{2}0]$ to $[10\bar{1}0]$ direction reveal the existence of an additional frequency.
 
 The dHvA signal as a function of magnetic field is presented in Fig.~\ref{Fig3_ok}(b) for $H \parallel$~[10$\bar{1}$0] and $T=25$~mK. The corresponding Fast Fourier Transform (FFT) spectrum exhibits \textit{two} peaks, as shown in Fig.~\ref{Fig3_ok}(c). The higher frequency of 690~T had been already observed and is referred to as the $\alpha$ branch \cite{Inada1999}. The lower frequency of 490~T, that we will refer to as $\alpha'$, is observed for the first time, with a much smaller amplitude compared to the $\alpha$ branch. The FFT spectrum of the dHvA signal taken for $H \parallel$~[11$\bar{2}$0] [see Fig.~\ref{Fig3_ok}(d)] clearly displays the lack of the $\alpha'$ branch peak in this direction. While the $\alpha$ branch is detected at all measured angles and does not change as previously reported \cite{Inada1999}, the $\alpha'$ branch is slightly shifting to higher frequencies and then disappears around $16\degree$ towards the [11$\bar{2}$0] direction.
 
 The effective masses were extracted for both frequencies, using systematic data at several temperatures and Lifshitz-Kosevitch fitting, for field along [10$\bar{1}$0]  and at $10\degree$ from [10$\bar{1}$0] to [11$\bar{2}$0].  For the $\alpha$ branch, the effective mass is almost angle-independent, yielding 12.1~$m_0$ for [10$\bar{1}$0] and 11.8~$m_0$ for $10\degree$, in accordance with previous measurements \cite{Inada1999}. The mass of the $\alpha'$ branch, on the other hand, changes with angle. It decreases from 24.6~$m_0$ at [10$\bar{1}$0] to 19.6~$m_0$ at $16\degree$.

We note that the observed angular dependence in the basal plane from [11$\bar{2}$0] to [10$\bar{1}$0] attributed to the $\alpha'$ branch is in good agreement with the calculated angular dependence of the $\alpha$ branch \cite{Inada1999}. 
 The $\eta$ branch is also supposed to exist in this angular range, but it is not observed experimentally. 
 In Ref. \onlinecite{Knopfle1996}, the calculated $\eta$ branch in the [10$\bar{1}$0] direction is situated at a frequency just below the one of the $\alpha$ branch, and may correspond to the observed $\alpha'$. Further refinement of band structure calculations may resolve this frequency mismatch.
 
 Altogether, the dHvA experiments below $H_M$ in the AF state do not reveal any sign of an electronic topological transition - we will come back to this fact later on. 


\subsection{Fermi surface above $H_M$}

\begin{figure}[h!]
	\begin{center}
		\includegraphics[width=8cm]{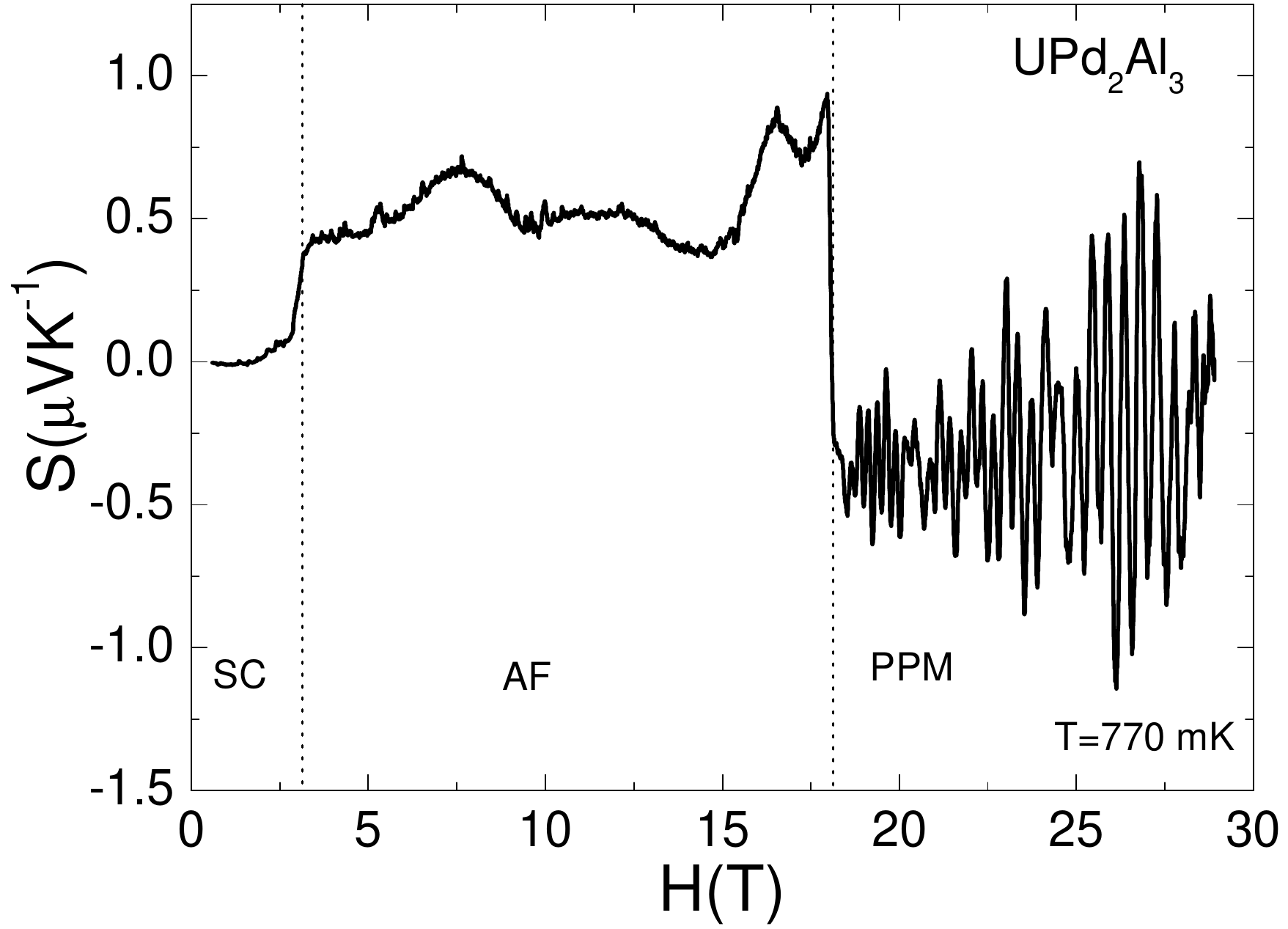}
		\caption{\label{Fig4_ok} Field dependence $S(H)$ of the TEP with thermal gradient along [11$\bar{2}$0] at $T=770$~mK for  H$\parallel$~[10$\bar{1}$0]. Quantum oscillations suddenly appear above a steep jump and sign change at the first order transition at $H_M$. Below $H_M$, there are five successive anomalies, but quantum oscillations are not observed.}
	\end{center}
\end{figure}

Figure~\ref{Fig4_ok} shows the field dependence of the TEP up to 28.9~T at $T= 770$~mK. Below $H_M$, five successive anomalies can be distinguished (which will be discussed in more detail later on). At $H_M=18$~T, coinciding with the first order metamagnetic transition, the TEP exhibits a sharp transition and jumps from more than $+0.5$~$\mu$V/K to about $-0.5$~$\mu$V/K (reminiscent of the sign change of the Hall effect at $H_M$), indicating a strong modification of the Fermi surface. Most spectacular, prominent magnetic quantum oscillations appear above $H_M$. The sudden observation of quantum oscillations in the TEP above $H_M$ is a clear-cut signature of an abrupt Fermi surface change by the appearance of either a new Fermi surface pocket due to the unfolding of the electronic bands or due to the sudden weakening of the renormalized effective mass due to the change of the electronic correlations resulting in a modification of the band structure. 

\begin{figure}[h!]
	\begin{center}
		\includegraphics[width=8cm]{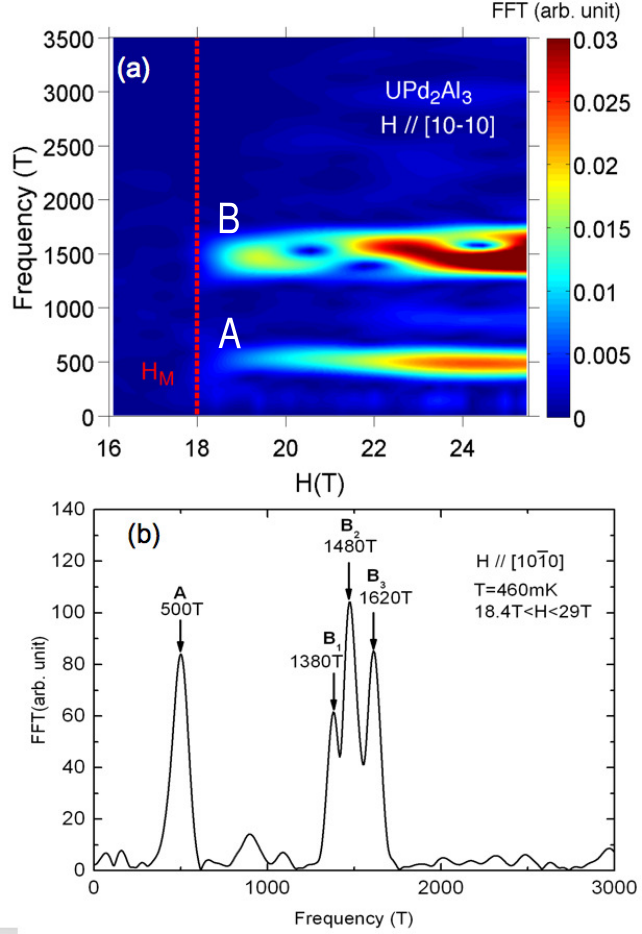}
		\caption{\label{Fig5_ok} (Color online) (a) Linear color map of the FFT intensity in the frequency-magnetic field plane from 15~T to 29~T for $H \parallel$ [10$\bar{1}$0]. The FFT analysis is done over a sliding window of width $\frac{1}{H_{\rm min}}-\frac{1}{H_{\rm max}}=0.01$~T$^{-1}$. (b) FFT of the TEP signal between 18.4~T and 28.9~T. A frequency of 500~T ($A$) and three close ones of 1380 ($B_1$) , 1480 ($B_2$) and 1620~T ($B_3$) are resolved.}
	\end{center}
\end{figure}

To study more carefully this abrupt appearance of quantum oscillations, we performed a FFT analysis over a sliding window of width $1/H_{\rm min}-1/H_{\rm max}=0.01~$T$^{-1}$ in the field range from 15~T to 29~T. The FFT intensity is represented as a linear color map in the frequency-magnetic field plane [see Fig.~\ref{Fig5_ok}(a)]. Note that the effective field $1/H_{\rm eff} = 1/2(1/H_{\rm min} + 1/H_{\rm max})$ is limited to the field range of 16~T to 25.5~T. Two main frequencies appear above $H_M=18$~T (represented by a red dashed line): $A$ at 500~T and $B$ around 1500~T. The evolution of the last one with field is very peculiar, with a splitting of the FFT peak into two ones with different spectral weight as a function of magnetic field. In fact, the FFT of the TEP signal over a large magnetic field range (between 18.4 and 29~T), shown in Fig.~\ref{Fig5_ok}(b), reveals three peaks at $B_1=1380$, $B_2=1480$ and $B_3=1620$~T. This indicates that a large magnetic field window is needed to separate three very close frequencies.

The Pantsulaya-Varlamov formula has been used to extract the effective masses from the temperature dependence of the amplitude of the TEP quantum oscillations \cite{Pantsulaya1989, PalacioMorales2016}. All these branches show small effective masses of 4.8~$m_0$ for $A$ and 4.6, 4.7 and 4.7~$m_0$ for the $B_1$, $B_2$ and $B_3$ frequencies, respectively. $A$ could correspond to the $\alpha'$ pocket detected in dHvA at lower field and lower temperature in the AF state (see Fig.~\ref{Fig3_ok}). This would imply a strong decrease of the effective mass, from 24.6~$m_0$ below $H_M$ to 4.8~$m_0$ above $H_M$, indicating a strong weakening of the correlations.

The three frequencies around 1500~T have not been detected below $H_M$ in the AF state in our TEP experiment. If we compare to existing band structure calculations below the metamagnetic transition, these frequencies could correspond to the $\delta$ and $\eta$ branches predicted in the AF state (see Fig.~\ref{Fig3_ok}(a) and \cite{Inada1999}). However, in the dHvA experiment of Ref.~\cite{Inada1999} no Fermi surface branch could be followed right up to the [10$\bar{1}$0] direction except the $\alpha$ branch, and it has been suggested that the unfavorable curvature factor (in particular of the $\delta$ branch) is responsible for this.

Definitely, in our TEP experiment we could not detect any Fermi surface branch below $H_M$. The most reasonable explanation for this is that the effective masses are too high below $H_M$ making their detection difficult in the temperature range of our experiment above 460~mK. The fact that we do observe quantum oscillations above $H_M$ strongly supports a drastic Fermi surface reconstruction at $H_M$ between the AF and the PPM state. 
In Ref.~\onlinecite{Terashima1997}, dHvA measurements have been performed through and above $H_M$ with the magnetic field along the $[10\bar{1}0]$ direction. The FFT of the oscillating signal for magnetic field from $H$ = 18.1 to 19.7~T shows two frequencies, one called $\Lambda$ at 1270~T and another called $\Xi$ at 3000~T. The effective masses for $\Lambda$ and $\Xi$ are 5.4~$m_0$ and 31~$m_0$, respectively. It is unlikely that the $\Lambda$ branch could correspond to the 1380~T observed here, besides the fact that the masses are very similar, the difference in the observed frequency is too large. The temperature limitation of the TEP experiment in high field above 460~mK can easily explain why the $\Xi$ pocket with a heavy mass of 31~$m_0$ has not been observed in our study. 

\begin{figure}[h!]
	\begin{center}
		\includegraphics[width=8cm]{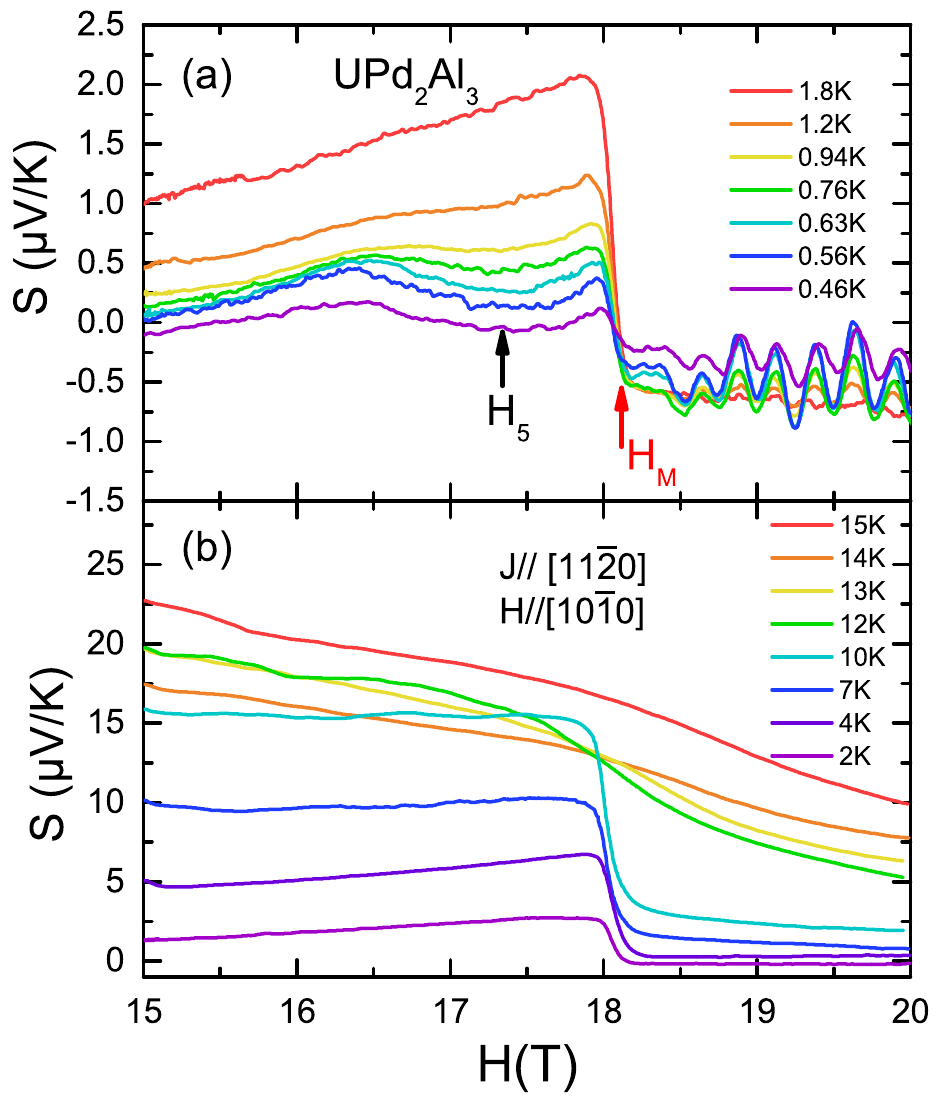}
		\caption{\label{Fig6_ok} (Color online). Temperature evolution of the TEP anomaly at $H_M$. (a) At low temperature, quantum oscillations appear above the first oder transition at $H_M$. The anomaly $H_5$ (black arrow, see text in section IV.) is also represented. (b) Above 10~K, the transitions broaden significantly, indicating a cross-over area between PM and PPM states.}
	\end{center}
\end{figure}

The anomaly at $H_M$ as seen by the TEP ($S(H)$) is presented in Fig. \ref{Fig6_ok} for several temperatures up to 15~K. Quantum oscillations above $H_M$ can be followed up to $T\approx 1$~K [see Fig.~\ref{Fig6_ok}(a)]. The step-like transition at $H_M$ can be followed up to $T\approx 10$~K and is very similar to what is observed in $S(H)$ at metamagnetic transitions of other compounds (UTe$_2$ \cite{Niu2020b}, UCoAl \cite{Palacio2013}, CeRh$_2$Si$_2$ \cite{Palacio2015}, CeRu$_2$Si$_2$ \cite{Pfau2012}). At higher temperatures, a broad cross-over occurs [see Fig.~\ref{Fig6_ok}(b)], in agreement with the previously reported phase diagram \cite{Sakon2002}. This suggests that the critical point of the first order transition is located close to 10~K.

\section{Successive Lifshitz transitions below $H_M$}

\subsection{Electronic topological transitions in thermoelectric power below $H_M$}

 \begin{figure}[h!]
 	\begin{center}
 		\includegraphics[width=8cm]{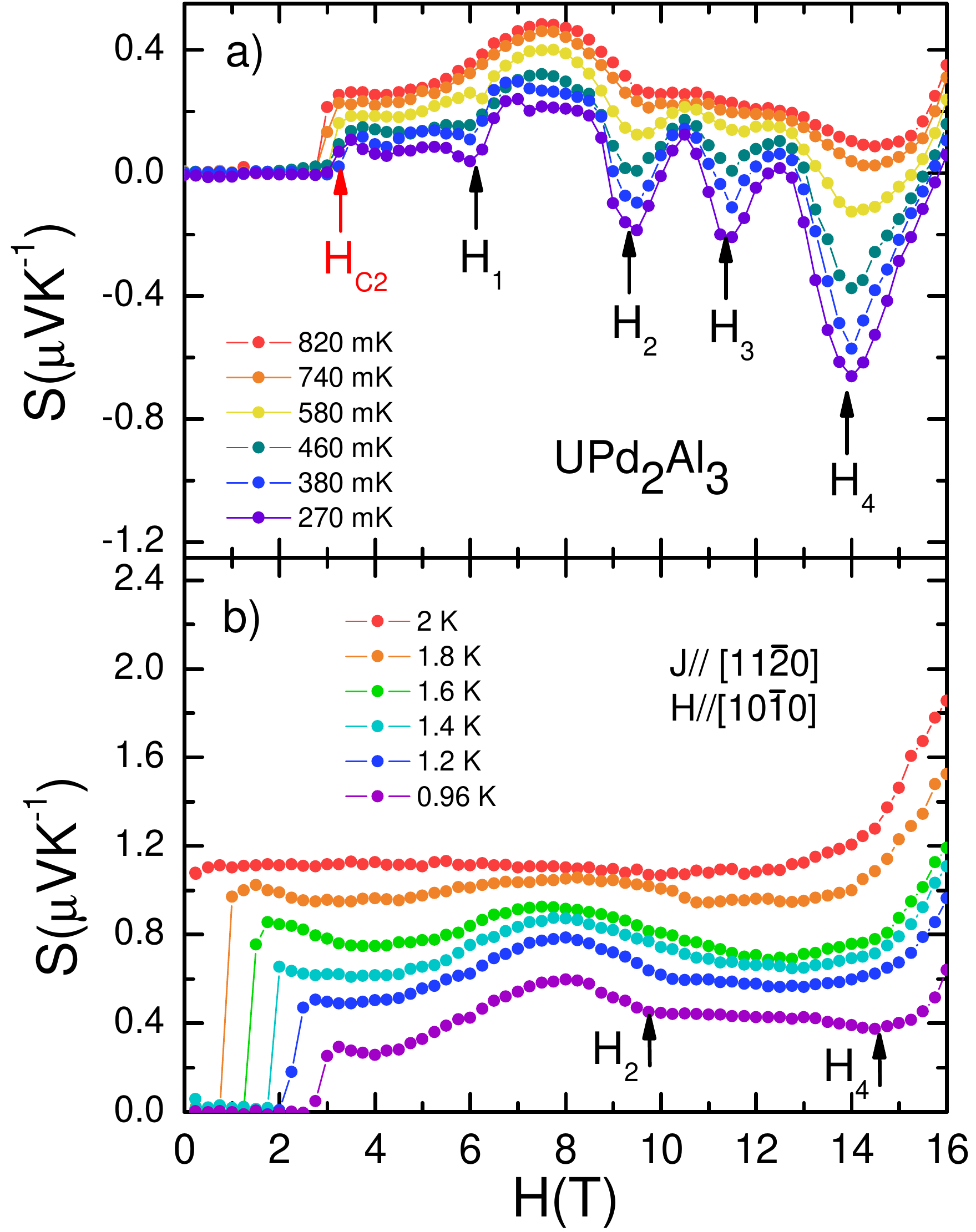}
 		\caption{\label{Fig7_ok} (Color online).  (a) A succession of anomalies (black vertical arrows) is observed at low temperatures in the thermoelectric response on a large magnetic field range between 4 and 16~T, for $H \parallel$[10$\bar{1}$0]. (b) Above 1~K, the different anomalies fade away and finally disappear.}
 	\end{center}
\end{figure}

Now we concentrate on the evolution of the TEP inside the AF state below $H_M$. For $H\parallel [10\bar{1}0]$ and at low temperatures, $S(H)$ is displayed in Fig.~\ref{Fig7_ok}(a). Above the superconducting critical field $H_{c2}\approx3.25~T$, at least five successive anomalies can be observed up to $H_M$. In the AF state the TEP signal starts with a positive sign and shows a first negative peak at $H_1\approx 6$~T, which is followed by a sharp maximum at lowest temperature. This maximum broadens rapidly with increasing temperature and for $T>0.5$~K a broad bump centered around $H\approx 7.5$~T develops. Upon increasing further the magnetic field, $S(H)$ exhibits three successive prominent minima at $H_2 \approx 9.5$~T, $H_3 \approx 11.5$~T and $H_4 \approx 14$~T. An additional minimum at $H_5\approx 17.5$~T has been observed, just before reaching the metamagnetic transition [see black arrow in Fig.~\ref{Fig6_ok}(a)]. With increasing temperature, these anomalies become less pronounced (while their field positions barely change), and finally they vanish above $T\approx1$~K, see Fig.~\ref{Fig7_ok}. The only remaining anomaly is the first order transition at $H_M$, turning into a cross-over regime above the critical point near 10~K, see Fig.~\ref{Fig6_ok}. 

Marked successive anomalies in $S(H)$ combined with the absence of any significant signature of possible phase transitions in thermodynamic properties such as magnetization \cite{Sugiyama2000} or of changes of the magnetic structure in neutron diffraction experiments \cite{Paolasini1994} suggests that the observed anomalies in $S(H)$ are related to topological Fermi surface changes. Similar successive anomalies in the TEP have been observed  in YbNi$_4$P$_2$ \cite{Pfau2017}, YbRh$_2$Si$_2$ \cite{Pfau2013, Pourret2013_2, Pourret2019} and UCoGe \cite{Bastien2016}. TEP is an extremely sensitive probe for such electronic topological transitions (like Lifshitz transitions) as it is sensitive not only to the changes of the DOS, but, more critically, also to the resulting changes in the scattering time which is the dominant contribution to the anomalies at Lifshitz transitions \cite{Varlamov1989}. A Lifshitz transition occurs when the Fermi level passes through a van Hove singularity in the electronic band structure, leading to a sharp feature (kink) in the DOS as a function of an external parameter, here the magnetic field. A huge TEP appears at such an electronic topological transition due to additional scattering processes from the main Fermi surface to the critical Fermi surface parts which are characterized as void formation (or equivalent to a neck disruption) or disappearance of a void (neck formation) \cite{Varlamov1989}. 

\begin{figure}[t!]
 	\begin{center}
 		\includegraphics[width=8cm]{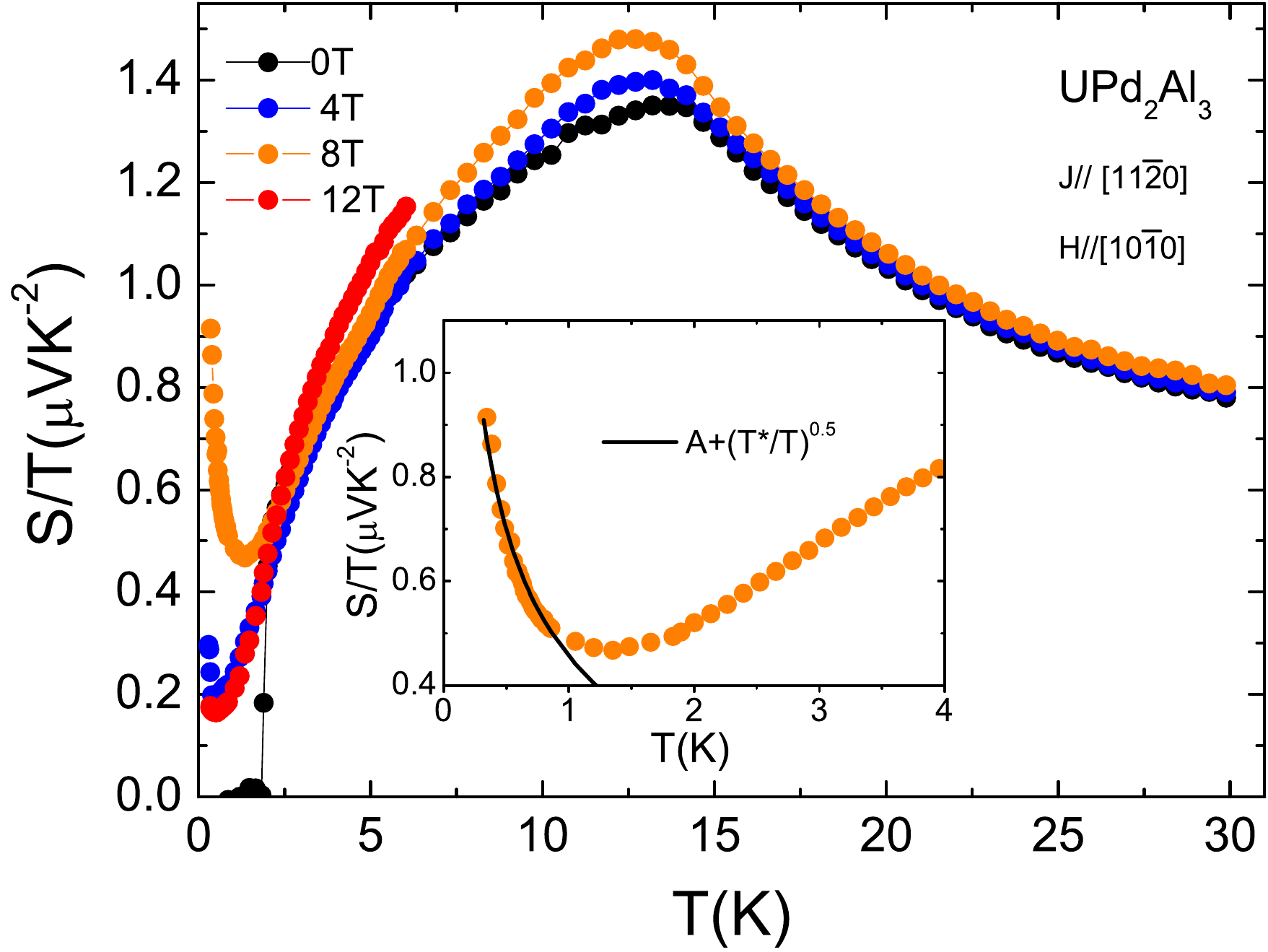}
 		\caption{\label{Fig8_ok} (Color online). Temperature dependence $S(T)/T$ for different magnetic fields. $S(T)/T$ shows a maximum at the PM-AF transition, $T_{N}=14.5$~K. At low temperature,  $S(T)/T$ does not get constant, indicating a non-Fermi liquid behavior. Inset: The $(T^{*}/T)^{0.5}$-dependence (see text) at $H=8$~T is characteristic for the proximity to a Lifshitz transition.}
 	\end{center}
\end{figure}

An additional indication that the anomalies observed in $S(H)$ in the AF state below $H_M$ are related to Lifschitz transitions stems from the temperature dependence $S(T)$ for fields close to a critical field where such a transition occurs. Fig.~\ref{Fig8_ok} shows the temperature dependence of  $S(T)/T$ at different fields. Firstly, it is obvious that the simple Fermi liquid regime $S(T)/T = {\rm const}$ for $T \to 0$ is never reached. The upturn in $S(T)/T$ at low temperature for $H=8~T$ is reminiscent of the proximity to a Lifshitz transition at $H_2$ and can be fitted (represented by a black line in the inset of Fig.~\ref{Fig8_ok}) by the expression $A+(T^*/T)^{0.5}$, which gives a characteristic energy $T^{*}\approx 350$~mK  \cite{Varlamov1989}. A similar behavior has been observed in UTe$_2$ where the characteristic temperature was about 150~mK \cite{Niu2020}.

\begin{figure}[h!]
	\begin{center}
		\includegraphics[width=8cm]{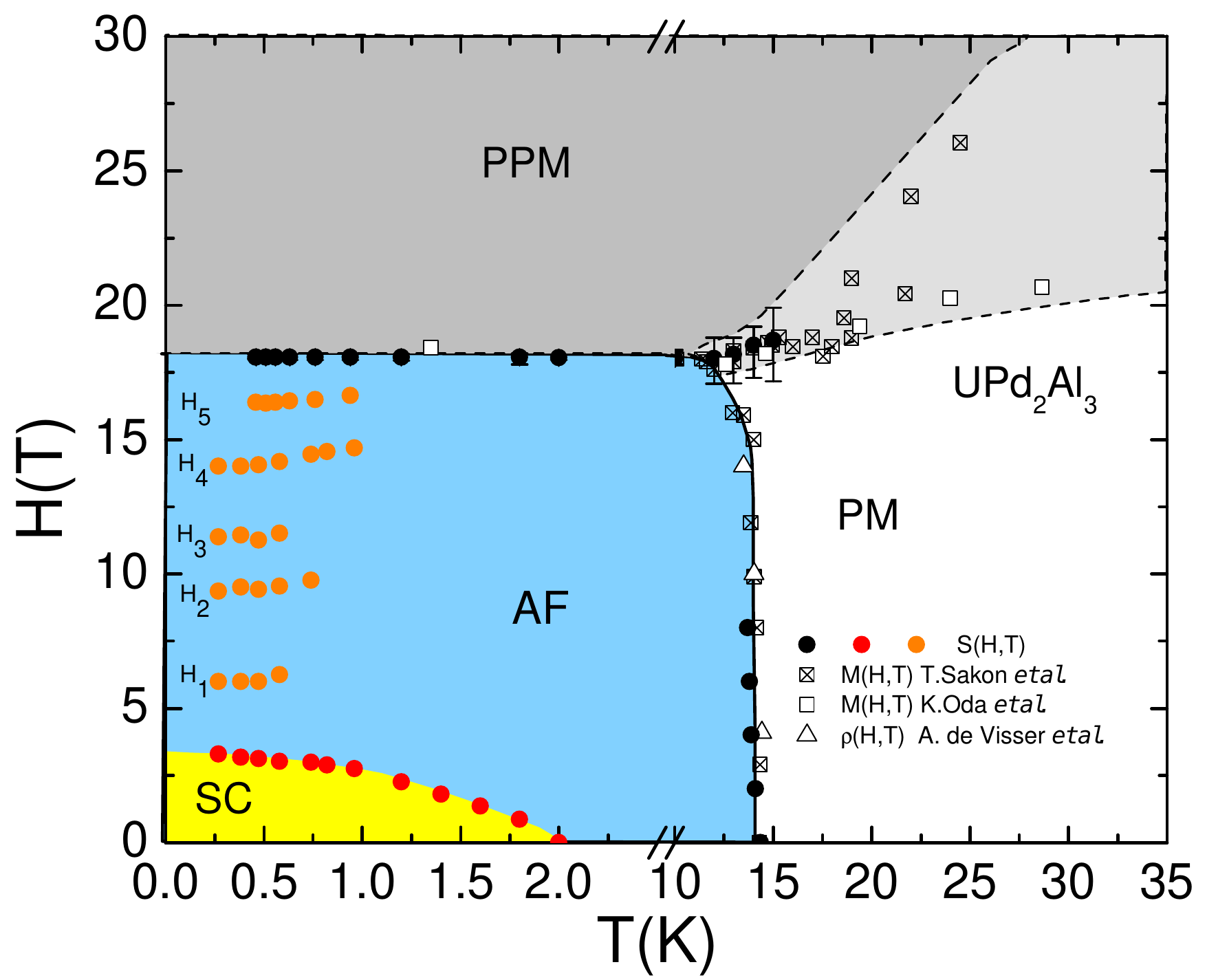}
		\caption{\label{Fig9_ok} (Color online). Phase diagram of UPd$_2$Al$_3$. SC denotes the superconducting state. Full circles have been extracted from the TEP measurements (the black vertical bars indicate the transition widths). Cross-squares and open squares represent magnetization data from Refs.~\onlinecite{Sakon2002} and \onlinecite{Oda1999} respectively, and open triangles magnetoresistance data \cite{DeVisser1994}. The light grey region corresponds to a cross-over regime between the PM and PPM states.}
	\end{center}
\end{figure}

The phase diagram containing the different TEP anomalies below $H_M$, and the anomalies at $H_M$ and $T_N$ is represented in Fig.~\ref{Fig9_ok}. The black vertical bars indicate the width of the corresponding transitions. Different anomalies, previously reported,  have been superimposed: Cross-squares and open squares are obtained by magnetization measurements from Refs.~\onlinecite{Sakon2002, Oda1999} respectively. Open triangles indicate transition points coming from magnetoresistance measurements \cite{DeVisser1994}. The light grey region corresponds to a cross-over regime between the PM and the PPM state, which is a common feature in many other heavy fermion compounds (URh$_2$Si$_2$, URhGe,...).

\color{black}

\subsection{Theoretical model}


It is well established that 5$f$-states in actinide intermetallics are not ordinary band states: standard DFT calculations fail to reproduce the narrow quasiparticle bands near $E_F$, while predicted 5$f$-bandwidths are too small to match photoemission data \cite{Allen92, Fujimori99}. This reflects the inadequate treatment of local (Hund's rule) correlations, which drive ``partial'' orbital-selective localization and high effective masses \cite{Zwicknagl2003}: some 5$f$-orbitals form extended Bloch states while others remain localized as multiplets. This dual character was first conjectured for UPd$_2$Al$_3$ \cite{Bohm92,Grauel92,Krimmel1992}, and direct evidence for coexisting 5$f$ quasiparticles and local magnetic excitations comes from neutron scattering \cite{Metoki1998, Bernhoeft1998, Hiess2006}. Localized 5$f$ states are even believed responsible for the pairing interaction in superconductivity (see reviews \onlinecite{Thalmeier2005a,Thalmeier2005b, Fulde2006, Zwicknagl2016}).

The dual model is an effective low-energy Hamiltonian (valid for $\hbar \omega \lesssim 10$~meV), obtained by integrating out high-energy processes: intra-atomic Hund's rule correlations renormalize the hybridization to zero for some channels while leaving others finite. The renormalized quasiparticles are itinerant 5$f$ electrons whose masses are dressed by low-energy excitations of the localized 5$f$ states. The scheme (Refs.~\onlinecite{Thalmeier2005a,Thalmeier2005b,Fulde2006,Zwicknagl2016,Petit2003,Wills2004,Zwicknagl2003a}) proceeds in three steps: (i) band-structure calculation of the bare itinerant 5$f$ dispersion; (ii) quantum-chemical calculation of the localized multiplets and their coupling to the itinerant states; (iii) many-body perturbation theory for the renormalized effective mass. All 5$f$ electrons are treated as fermions throughout.

Conventional LDA calculations treating all 5$f$-states as itinerant reproduce ground-state properties (Fermi surface, densities), but this is not conclusive evidence of itinerant character: localized states can mimic filled bands far below $E_F$, and the Fermi surface --- fixed mainly by particle number --- is insensitive to changes by an even number of electrons (e.g., a filled band). Unambiguous proof of the dual character instead comes from spectral functions, which show high-energy features from transitions into excited local multiplets \cite{Fujimori2012, Fujimori2014,Fujimori2016,Fujimori2019}.

\begin{table}[h!]
\caption{\label{Table1}Effective masses in UPd$_2$Al$_3$ for $H \parallel c$. Notation for Fermi surface sheets and experimental values
from Ref.~\onlinecite{Inada1999}. Theoretical values from Ref.~\onlinecite{Zwicknagl2003}}.
\begin{tabular}{ccc}
 \hline
 \hline
 Fermi surface sheet & $m^*/m_0$ (exp.) & $m^*/m_0$ (theory) \\
 \hline
$\zeta$ & 65 & 59.6\\
$\gamma$ & 33 & 31.9 \\
$\beta$ &  19 & 25.1\\
$\epsilon_2$ &  18 &17.4 \\
$\epsilon_3$ &  12 &13.4 \\
$\alpha$ &  5.7 & 9.6 \\
\hline
\hline
\end{tabular}
\end{table}

\begin{figure}[h!]
	\begin{center}
		\includegraphics[width=5cm]{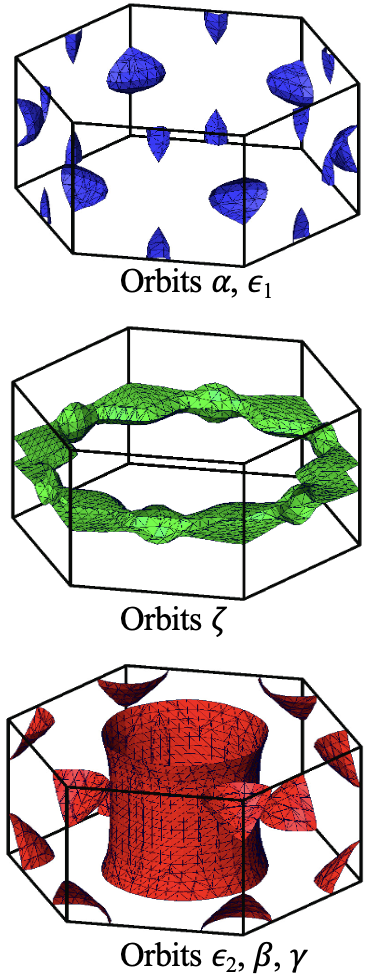}
		\caption{\label{Fig10_ok} (Color online) Fermi surface of UPd$_2$Al$_3$ calculated within the dual approach \cite{Zwicknagl2003} for $H=0$~T (assignment of the extremal orbits as in Ref.~\onlinecite{Inada1999}). The main cylinder part (red, $\beta$ and $\gamma$ sheets) has effective masses with $m^*= 19-33m_0$, the highest masses are found on the torus (green, $\zeta$ sheet).}
	\end{center}
\end{figure}

This scheme, applied to UPd$_2$Al$_3$ and UPt$_3$, gives parameter-free agreement with measured dHvA frequencies and effective masses (Table~\ref{Table1}). The bare 5$f$ dispersion is obtained by solving the Dirac equation self-consistently in LDA, excluding two U $5f$ ($j=5/2$) states from band formation \cite{Christensen84} --- motivated by the absence of Kramers' degeneracy, suggesting an even number of localized 5$f$ electrons. Since the heavy quasiparticles form only below $T_N \approx 14.5$~K, where localized moments order ($\approx 1\mu_B$), the calculation uses the observed AF structure, whose superstructure strongly affects the heavy quasi-particle Fermi surface.

\begin{figure}[h!]
	\begin{center}
		\includegraphics[width=\linewidth]{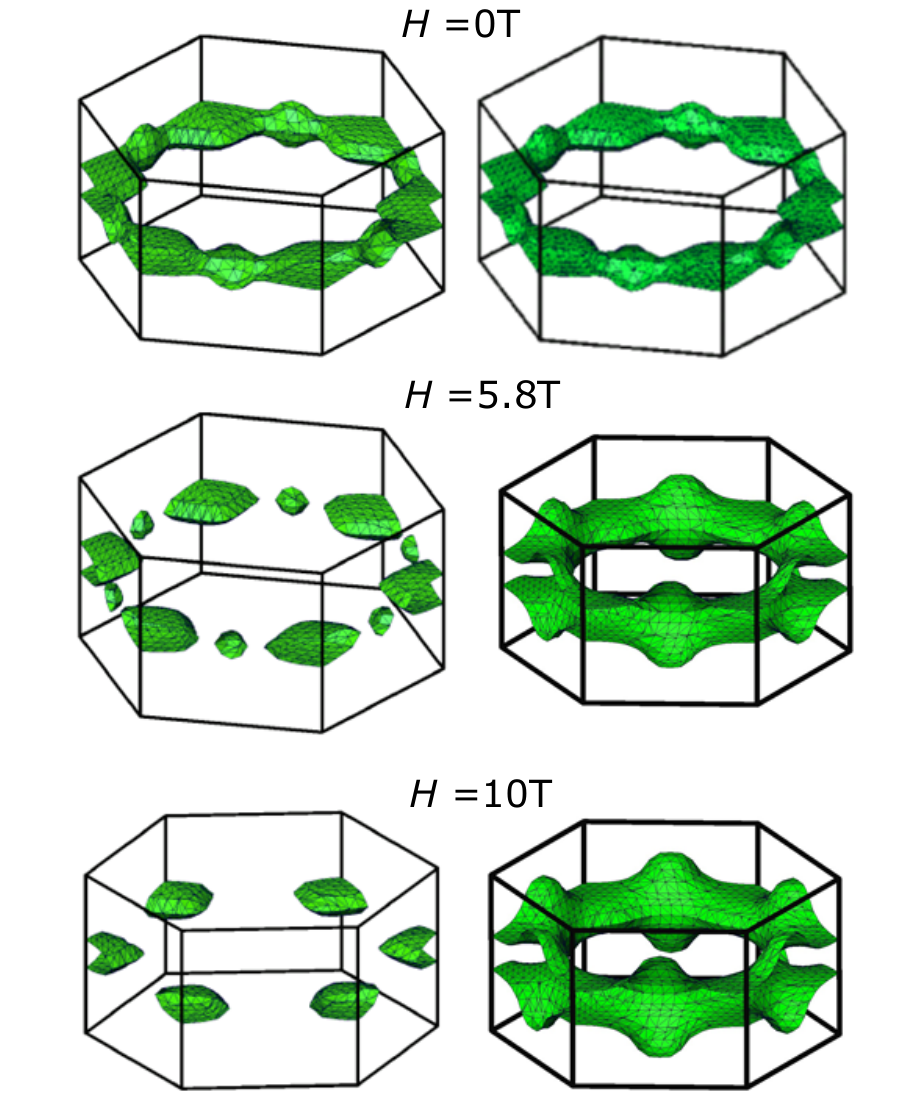}
		\caption{\label{Fig11_ok} (Color online) Left panels: Lifshitz transitions of the majority-spin band $\zeta$ Fermi surface in the AF state at $H_1\approx5.8$~T (neck breaking) and $H_2\approx10$~T (disappearance of small pocket). Right panels: The minority-spin band $\zeta$ Fermi surface is just increasing under magnetic field without any Lifshitz transition.}
	\end{center}
\end{figure}

In the second step, the localized U $5f$ states follow from diagonalizing the Coulomb matrix (evaluated with $jj$-coupled radial functions from the ab-initio potential) in the localized subspace; the coupling to delocalized $5f$ electrons comes from Coulomb matrix elements in the $5f^3$ configuration. The mass renormalization then follows from scattering of itinerant $5f$ electrons off low-energy excitations of the localized $5f^2$ multiplet --- analogous to the factor-5 mass enhancement from CEF (crystal electric field) excitations in Pr metal \cite{Fulde81}. The effective masses in Table \ref{Table1} are obtained from an isotropical renormalization of the band mass $m_b$ given by: $\frac{m^*}{m_b}=1-\frac{\partial \Sigma(\omega)}{\partial \omega}|_{\omega=0}$. The explicit expression of the local self-energy of the delocalized $5f$ states $\Sigma(\omega)$ is given in Ref.~\onlinecite{Zwicknagl2003a}. The mass enhancement is calculated self-consistently inserting values for the density of states at the Fermi level $N(0)=2.76$~states/(eV cell spin) obtained from the band-structure, when two $5f$ electrons are kept localized. The vertex is given by $a|M|= 0.084$~eV where the prefactor $a$ denotes the 5$f$ weight per spin and U atom of the conduction electron states near $E_F$. The matrix element $M$ describes the transition between the localized states $|\Gamma_4⟩$ and $|\Gamma_3⟩$ due to the Coulomb interaction $U_\text{Coul}$ with the delocalized 5$f$ electrons. Finally, the dynamical susceptibility is approximated by that of an eﬀective two-level system with an excitation energy $\delta\approx7$~meV.

\begin{figure}[h!]
	\begin{center}
		\includegraphics[width=8cm]{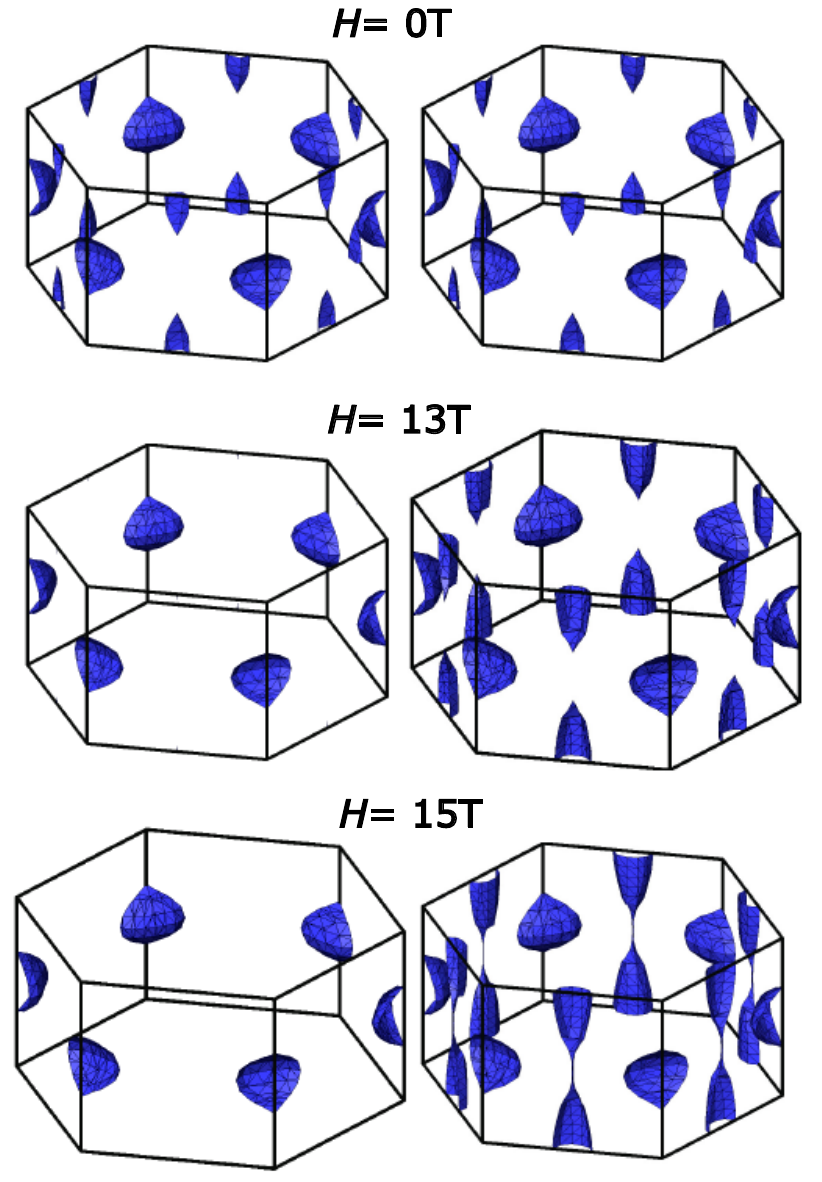}
		\caption{\label{Fig12_ok} (Color online) Lifshitz transitions of the $\epsilon_1$ Fermi surface pockets in the AF state (the $\alpha$ pockets, situated in the middle of the six edges of the  hexagonal prism, are not concerned by the Lifshitz transitions). Left panels: Majority-spin band Fermi surface with Lifshitz transition at $H_3\approx13$~T: disappearance of the small $\epsilon_1$ pocket. Right panels: Minority-spin band Fermi surface with Lifshitz transition at $H_4\approx15$~T: neck formation in the $\epsilon_1$ pocket.}
	\end{center}
\end{figure}

The Fermi surface in the AF state, shown in Fig.~\ref{Fig10_ok}, is composed of three Kramers-degenerate bands: the $\alpha$ (hole), $\beta$ (electron), $\gamma$ (electron), $\zeta$ (hole), $\epsilon_1$ (hole), and $\epsilon_2$ (hole) sheets. An external magnetic field lifts this degeneracy via the Zeeman effect. Within a rigid-band approximation, we calculate the corresponding Fermi surfaces for the majority- and minority-spin bands. Using a $g$-factor of 1.9 for magnetic fields applied in the basal plane, we identify four Lifshitz transitions occurring at $H=5.8$~T (neck breaking in the $\zeta$ majority-spin band, see Fig.~\ref{Fig11_ok}), 10~T (disappearance of a small pocket associated with the $\zeta$ majority-spin band, see Fig.~\ref{Fig11_ok}), 13~T (disappearance of the small $\epsilon_1$ pocket of the majority-spin band, see Fig.~\ref{Fig12_ok}), and 15~T (neck formation in the $\epsilon_1$ sheet of the minority-spin band, see Fig.~\ref{Fig12_ok}). This predicted sequence of field-induced topological changes agrees well with the successive anomalies observed in the field dependence of the TEP, $S(H)$, see Fig.~\ref{Fig6_ok}.

Because both the type of topological change (e.g., void formation, neck breaking) and the carrier character (electron or hole) are known \cite{Pfau2017}, one can proceed with a more refined spectroscopic analysis of the thermoelectric signal, following the approach of the seminal study on YRh$_2$Si$_2$ \cite{Pourret2019}. The key principle of the theoretical framework is that for a Lifshitz transition the scattering term (appearance of a new scattering channel) plays a far more decisive role in determining transport coefficients than changes in the DOS. For instance, when a new pocket is created in the Brillouin zone (void formation), it can act as a trap for electrons or holes in the low-temperature impurity-scattering regime \cite{Varlamov2021}. Although this has only a modest impact on the total DOS, it crucially modifies the energy dependence of the scattering time, which can increase substantially. The resulting thermoelectric response typically has the shape of an asymmetric peak appearing slightly above or below the critical energy, depending on whether a void or a neck is created or destroyed; the precise peak position also exhibits a weak temperature dependence. However, as demonstrated in Ref.~\onlinecite{Pourret2019}, when several topological transitions occur in close succession, the associated scattering processes can no longer be treated independently, and the thermoelectric response becomes correspondingly more complex.
All successive anomalies in $S(H)$ appear as negative peaks, consistent with the hole character of the $\zeta$ and $\epsilon_1$ pockets involved. Because the magnetoresistance does not display corresponding anomalies, determining the precise nature of each topological change is not straightforward.  In general, the magnetoresistance is less sensitive to the appearance or disappearance of new pockets and dominated by the large Fermi surfaces. Nevertheless, the anomalies in $S(H)$ are observed at fields slightly below our calculated critical fields for $H_2$, $H_3$, and $H_4$ which is expected for Lifshitz transitions at $T\neq 0$, see Ref.~\onlinecite{Varlamov1989}. Furthermore, this is consistent with the disappearance of pockets at $H_2$ and $H_3$ and with neck formation at $H_4$. In contrast, the anomaly at $H_1$ appears at a slightly higher field than predicted, which is compatible with a neck-breaking transition.
These considerations demonstrate that the thermoelectric anomalies arise naturally from the sequence of field-induced Lifshitz transitions and that the detailed comparison between theory and experiment provides a consistent picture of the evolution of the Fermi surface in the AF state.

\section{Discussion}

The field induced Lifshitz transitions reported here are very similar to what has been reported in YbNi$_4$P$_2$ \cite{Pfau2017} with nine successive anomalies and in YbRh$_2$Si$_2$ \cite{Pfau2013,Pourret2013_2,Pourret2019} with three anomalies. For U systems we can mention  the ferromagnetic superconductor UCoGe, where up to five anomalies have been observed in different transport quantities for a field along the $c$ axis \cite{Bastien2016, Leenen2024}. Another very similar system appears to be URu$_2$Si$_2$. In the so-called hidden order phase, under magnetic field, four anomalies have been reported in different transport properties such as TEP, Nernst signal, resistivity and Hall resistivity measurements. The field positions of these anomalies correspond to different changes in the Shubnikov-de Haas frequencies and effective masses around $H_k=12$, $H_p=17$, $H^*=23$, and $H_r=30$~T \cite{Shishido2009, Aoki2012_URu2Si2, palacio2013b}. These results indicate successive Fermi surface instabilities inside the hidden order phase induced by magnetic field.

In all three U-based systems, a distinct modification of the Fermi surface occurs when the magnetic polarization of the 5$f$ electrons reaches a critical value driven by the Zeeman effect. The rudimentary picture is that due to the Zeeman effect the cross section of either the majority or minority-spin band decreases or increases leading to an electronic instability and thus driving the electronic topological transition. Surprisingly, the value of the field-induced critical polarizations are very similar in these three U-based heavy fermion systems, see Tab.~\ref{Table2}. For UPd$_2$Al$_3$ the sequence of observed anomalies corresponds to the theoretical predictions of the dual model approach (orbital-selective localization) and definitely validates the dual character framework (localized and delocalized $5f$-orbitals). The shape of the anomalies are in good agreement with the sign of the charge carriers and with the proposed topological changes. Nevertheless, the precise nature of each Lifshitz transition below $H_M$ (e.g. void formation vs. neck disruption) is difficult to determine solely from TEP, as magnetoresistance measurements do not show corresponding anomalies. Investigating the temperature dependence of the anomalies more closely may contribute to fully characterize the critical fields and the corresponding phase boundaries.

We also stress that in the case of UCoGe the Lifshitz transitions have been clearly observed in quantum oscillation experiments, especially by the vanishing of one orbit with increasing field \cite{Bastien2016, Leenen2024}. In quantum oscillation measurements in URu$_2$Si$_2$ new orbits appear under magnetic field in the hidden order phase \cite{Shishido2009, Aoki2012_URu2Si2, Scheerer2014a}. Yet, in the AF state of UPd$_2$Al$_3$, the existing dHvA  results do not evidence any change of the Fermi surface below $H_M$. While UCoGe and URu$_2$Si$_2$ are low carrier systems with small Fermi surfaces, this is not the case in UPd$_2$Al$_3$, where the topological transitions occur only on either the minority-spin or the majority-spin Fermi surface, while the other bands remain unaffected, respectively. More specifically in UPd$_2$Al$_3$, looking at the $\zeta$ Fermi surface sheet below $H_M$ in the AF state, the corresponding frequency is observed by dHvA experiments up to the maximum field of 17~T (see Fig.~\ref{Fig3_ok} and \cite{Inada1999}), whereas two Lifshitz transitions occur (see Fig.~\ref{Fig11_ok}), at $H_1\approx5.8$~T and $H_2\approx10$~T. This apparent contradiction is removed when considering that both Lifshitz transitions only concern the $\zeta$ majority-spin band, whereas the minority one is not affected, which overall may imply only slight, not easy to detect changes in the oscillation amplitude, but not a complete disappearance nor an evident frequency shift. The situation is different for the $\epsilon_1$ Fermi surface pocket, which involves both spin bands (considering the successive Lifshitz transitions at $H_3\approx13$~T and $H_4\approx15$~T, see Fig.~\ref{Fig12_ok}). Hence one would expect the corresponding quantum oscillations being more significantly affected, at least at one of the two transitions. As a matter of fact, the $\epsilon_1$ orbits have not yet been detected by any dHvA experiment, preventing a more detailed comparison with theory.

The sharp transition at the metamagnetic field $H_M=18$~T appears as a drastic reconstruction of the Fermi surface. The sudden change of sign observed simultaneously in both the TEP ($S$) and the Hall coefficient ($\rho_{xy}$), from positive to negative, provides direct experimental evidence for the change of the dominant charge carriers, from hole-like in the AF state to electron-like in the PPM state. The corresponding abrupt appearance of substantial quantum oscillations above $H_M$ 
points to a major change in the electronic structure. Crucially, the resulting light effective masses, compared to heavier ones below $H_M$, suggest a strong weakening of electronic correlations and band unfolding in the PPM state. The higher effective masses below $H_M$ make the detection of quantum oscillations difficult or impossible in the TEP experiment above 460~mK. Further detailed studies (e.g. dHvA measurements at lower temperatures or higher fields) are needed to precisely map out the entire Fermi surface in the PPM state and confirm the effective masses and new pockets observed via TEP. 

Concerning the observed anomaly at $H_5\approx 17.5$~T, it might also be related to a Fermi surface instability and be part of the cascade along with the other four anomalies, but we lack data at temperatures below 460~mK to push the analysis further and unambiguously elucidate its origin (which is left for further studies).

\begin{table}[h!]
\caption{\label{Table2}List of magnetic fields and the corresponding polarization at which FS instabilities occur in UPd$_2$Al$_3$, UCoGe and URu$_2$Si$_2$\\}
\begin{tabular}{ccc}
  \hline
  \hline
 
\multicolumn {1}{c}{ compound} &\multicolumn {1}{c}{critical magnetic} & \multicolumn{1}{c}{field induced } \\
\multicolumn {1}{c} {} &\multicolumn {1}{c}{field (T)} & \multicolumn{1}{c}{polarization ($\mu_B/U$)} \\
 \hline
 UPd$_2$Al$_3$& H$_1$=6~T & 0.105\\
                            & H$_2$=9.5~T & 0.180\\   
  			 & H$_3$=11.5~T & 0.215 \\
			&  H$_4$=14~T & 0.275\\
			&  H$_5$=16.5~T & 0.340 \\ 
			
\hline

UCoGe & H$_1$=3.6~T \cite{Bastien2016} & 0.15 \cite{Knafo2012}\\
             &H$_2$=9.2~T & 0.27\\   
  	     & H$_3$=13~T & 0.36 \\
			&H$_4$=16~T & 0.41 \\
			&H$_5$=21~T & 0.5\\
\hline			
 URu$_2$Si$_2$& H$_k$=8.5~T \cite{palacio2013b} & 0.09  \\
                            & H$_p$=16~T & 0.160\\   
  			& H$^*$=23~T & 0.235 \\
			&H$_r$=30~T & 0.315 \\
    \hline
    \hline
\end{tabular}
\end{table}

%

\section{Conclusion}
Through thorough TEP measurements, we have unveiled a cascade of at least four successive field-induced Lifshitz transitions within the AF state of UPd$_2$Al$_3$. At $H_M$, across the first-order metamagnetic transition, we provide compelling evidence for a drastic Fermi surface reconstruction by a change of sign of the dominant charge carriers and a strong weakening of the electronic correlations (a reduction of the effective mass above $H_M$). Hence this work confirms that field-induced topological changes of the Fermi surface are a general feature of the phase diagrams of U-based heavy fermion systems. The sequence of successive field-induced Lifshitz transitions observed here is highly similar to what has been reported in other systems like the ferromagnetic superconductor UCoGe or the hidden order compound URu$_2$Si$_2$, suggesting a common underlying physics related to the critical magnetic polarization of $5f$-electrons. The quantitative agreement between the experimentally observed Lifshitz transitions and the theoretical predictions from the renormalized band structure calculations, taking into account the dual-nature of the 5$f$ states supports this approach as a robust framework for understanding orbital-selective localization in actinide heavy fermion materials. Furtheron, this study highlights the extreme sensitivity of the Fermi surface topology to magnetic field tuning in UPd$_2$Al$_3$ and reinforces the role of topological changes as a fundamental driving force in correlated electron systems. 
\vspace{0.25cm}

\acknowledgements
We thank H.~Harima for fruitful discussions. We acknowledge support of the LNCMI-CNRS, member of the European Magnetic Field Laboratory (EMFL), and from the Laboratoire d’excellence LANEF (ANR-10-LABX-0051).

\bibliography{references}

@article{Lifshitz,
author = {I. M. Lifshitz},
title = {Anomalies of Electron Characteristics of a Metal in the High Pressure Region},
journal = {Sov. Phys. JETP},
volume = {11},
pages = {1130},
year = {1960},
note = {{Zh. Eksp. Teor. Fiz. \textbf{38}, 1569 (1960)}},
url = {https://www.jetp.ras.ru/cgi-bin/e/index/e/11/5/p1130?a=list}
}

@article{Metoki1998,
  title = {Superconducting Energy Gap Observed in the Magnetic Excitation Spectra of a Heavy Fermion Superconductor {${\mathrm{UPd}}_{2}{\mathrm{Al}}_{3}$}},
  author = {Metoki, Naoto and Haga, Yoshinori and Koike, Yoshihiro and \=Onuki, Yoshichika},
  journal = {Phys. Rev. Lett.},
  volume = {80},
  issue = {24},
  pages = {5417--5420},
  numpages = {0},
  year = {1998},
  month = {Jun},
  publisher = {American Physical Society},
  doi = {10.1103/PhysRevLett.80.5417},
  url = {https://link.aps.org/doi/10.1103/PhysRevLett.80.5417}
}

@article{Shishido2009,
author = {Shishido, H. and Hashimoto, K. and Shibauchi, T. and Sasaki, T. and Oizumi, H. and Kobayashi, N. and Takamasu, T. and Takehana, K. and Imanaka, Y. and Matsuda, T. and Haga, Y. and Onuki, Y. and Matsuda, Y.},
doi = {10.1103/PhysRevLett.102.156403},
issn = {0031-9007},
journal = {Phys. Rev. Lett.},
month = {apr},
number = {15},
pages = {156403},
title = {{Possible Phase Transition Deep Inside the Hidden Order Phase of Ultraclean {URu$_2$Si$_2$}}},
url = {http://link.aps.org/doi/10.1103/PhysRevLett.102.156403},
volume = {102},
year = {2009}
}

@article{Bernhoeft1998,
  title = {Enhancement of Magnetic Fluctuations on Passing below ${T}_{c}$ in the Heavy Fermion Superconductor {${\mathrm{UPd}}_{2}{\mathrm{Al}}_{3}$}},
  author = {Bernhoeft, N. and Sato, N. and Roessli, B. and Aso, N. and Hiess, A. and Lander, G. H. and Endoh, Y. and Komatsubara, T.},
  journal = {Phys. Rev. Lett.},
  volume = {81},
  issue = {19},
  pages = {4244--4247},
  numpages = {0},
  year = {1998},
  month = {Nov},
  publisher = {American Physical Society},
  doi = {10.1103/PhysRevLett.81.4244},
  url = {https://link.aps.org/doi/10.1103/PhysRevLett.81.4244}
}

@article{Hiess2006,
Author = {Hiess, A. and Bernhoeft, N. and Metoki, N. and Lander, G. H. and
   Roessli, B. and Sato, N. K. and Aso, N. and Haga, Y. and Koike, Y. and
   Komatsubara, T. and Onuki, Y.},
Title = {Magnetisation dynamics in the normal and superconducting phases of
   {UPd$_2$Al$_3$}:: I.: Surveys in reciprocal space using
   neutron inelastic scattering},
Journal = {J. Phys.: Condens. Matter},
Year = {2006},
Volume = {18},
Number = {27},
Pages = {R437-R451},
Month = {JUL 12},
DOI = {10.1088/0953-8984/18/27/R01},
url = {https://doi.org/10.1088/0953-8984/18/27/R01},
ISSN = {0953-8984},
EISSN = {1361-648X},
ResearcherID-Numbers = {Komatsubara, Tetsuta/GSI-7012-2022
   Haga, Yoshinori/C-7679-2011
   Hiess, Arno/LZG-7589-2025},
ORCID-Numbers = {Haga, Yoshinori/0000-0002-4605-5117
   Hiess, Arno/0000-0002-6457-691X},
Unique-ID = {WOS:000238612900002},
}

@article{Yelland2011b,
author = {Yelland, E. A. and Barraclough, J. M. and Wang, W. and Kamenev, K. V. and Huxley, a. D.},
doi = {10.1038/nphys2073},
issn = {1745-2473},
journal = {Nature Physics},
month = {aug},
number = {11},
pages = {890--894},
publisher = {Nature Publishing Group},
title = {{High-field superconductivity at an electronic topological transition in URhGe}},
url = {http://www.nature.com/doifinder/10.1038/nphys2073},
volume = {7},
year = {2011}
}

@article{Blanter,
author = {Y. M. Blanter and M. I. Kaganov and A. V. Pantsulaya and A. A. Varlamov},
title = {The Theory of Electronic Topological Transitions},
journal = {Phys. Rep.},
volume = {245},
pages = {159},
year = {1994},
doi = {https://doi.org/10.1016/0370-1573(94)90103-1}
}

@article{Norman2010,
author = {M. R. Norman and J. Lin and A. J. Millis},
title = {Fermi-surface reconstruction and the origin of high-temperature superconductivity},
journal = {Phys. Rev. B},
volume = {81},
pages = {180513},
year = {2010},
doi = {https://doi.org/10.1103/PhysRevB.81.180513}
}

@article{Leboeuf2011,
author = {D. LeBoeuf and N. Doiron-Leyraud and B. Vignolle and M. Sutherland and B. J. Ramshaw and J. Levallois and R. Daou and F. Laliberte and O. Cyr-Choiniere and J. Chang and Y. J. Jo and L. Balicas and R. Liang and D. A. Bonn and W. N. Hardy and C. Proust and L. Taillefer},
title = {Lifshitz critical point in the cuprate superconductor {YBa$_2$Cu$_3$O$_y$} from high-field {Hall} effect measurements},
journal = {Phys. Rev. B},
volume = {83},
pages = {054506},
year = {2011},
doi = {https://doi.org/10.1103/PhysRevB.83.054506}
}

@article{Liu2010,
author = {C. Liu and T. Kondo and R. M. Fernandes and A. D. Palczewski and E. D. Mun and N. Ni and A. N. Thaler and A. Bostwick and E. Rotenberg and J. Schmalian and S. L. Bud'ko and P. C. Canfield and A. Kaminski},
title = {Electronic properties of iron-based superconductors},
journal = {Nat. Phys.},
volume = {6},
pages = {419},
year = {2010},
doi={https://doi.org/10.1038/nphys1656}
}

@article{Wang2015,
author = {Y. Wang and M. N. Gastiasoro and B. M. Andersen and M. Tomi and H. O. Jeschke and R. Valenti and I. Paul and P. J. Hirschfeld},
title = {Disorder effects in iron-based superconductors},
journal = {Phys. Rev. Lett.},
volume = {114},
pages = {097003},
year = {2015},
doi={https://doi.org/10.1103/PhysRevLett.114.097003}
}

@article{Pfau2013,
author = {H. Pfau and R. Daou and S. Lausberg and H. R. Naren and M. Brando and S. Friedemann and S. Wirth and T. Westerkamp and U. Stockert and P. Gegenwart and C. Krellner and C. Geibel and G. Zwicknagl and F. Steglich},
title = {Metamagnetism and Fermi surface effects in heavy fermion systems},
journal = {Phys. Rev. Lett.},
volume = {110},
pages = {256403},
year = {2013},
doi={https://doi.org/10.1103/PhysRevLett.110.256403}
}

@article{Pourret2013_2,
author = {A. Pourret and G. Knebel and T. D. Matsuda and G. Lapertot and J. Flouquet},
title = {{Magnetic polarization and Fermi surface instability: Case of YbRh$_2$Si$_2$}},
journal = {J. Phys. Soc. Jpn.},
volume = {82},
pages = {053704},
year = {2013},
doi = {10.7566/JPSJ.82.053704},
URL = {https://doi.org/10.7566/JPSJ.82.053704}
}

@article{Pfau2017,
  title = {{Cascade of Magnetic-Field-Induced Lifshitz Transitions in the Ferromagnetic Kondo Lattice Material YbNi$_{4}$P$_{2}$}},
  author = {Pfau, H. and Daou, R. and Friedemann, S. and Karbassi, S. and Ghannadzadeh, S. and K\"uchler, R. and Hamann, S. and Steppke, A. and Sun, D. and K\"onig, M. and Mackenzie, A. P. and Kliemt, K. and Krellner, C. and Brando, M.},
  journal = {Phys. Rev. Lett.},
  volume = {119},
  issue = {12},
  pages = {126402},
  numpages = {6},
  year = {2017},
  month = {Sep},
  publisher = {American Physical Society},
  doi = {10.1103/PhysRevLett.119.126402},
  url = {https://link.aps.org/doi/10.1103/PhysRevLett.119.126402},
}

@article{Kramers,
author = {H. A. Kramers},
title = {On magnetic resonance effects},
journal = {Le Magnetisme},
pages = {97},
year = {1940},
note = {See J. Bequerelle, Institut International de Cooperation Intellectuelle, CNRS}
}

@article{Stryjewski1977,
author = {E. Stryjewski and N. Giordano},
title = {Metamagnetism},
journal = {Advances in Physics},
volume = {26},
number = {5},
pages = {487--650},
year = {1977},
publisher = {Taylor \& Francis},
doi = {10.1080/00018737700101433},
URL = { https://doi.org/10.1080/00018737700101433
},
}

@article{Levitin1988,
author = {R. Z. Levetin and A. S. Markosyan},
title = {Magnetic properties of metals},
journal = {Usp. Fiz. Nauk.},
volume = {155},
pages = {623},
year = {1988},
note = {Sov. Phys. Usp. 31, 730 (1988)}
}

@article{Perry2001,
  title = {Metamagnetism and Critical Fluctuations in High Quality Single Crystals of the Bilayer Ruthenate {${\mathrm{Sr}}_{3}{\mathrm{Ru}}_{2}{\mathrm{O}}_{7}$}},
  author = {Perry, R. S. and Galvin, L. M. and Grigera, S. A. and Capogna, L. and Schofield, A. J. and Mackenzie, A. P. and Chiao, M. and Julian, S. R. and Ikeda, S. I. and Nakatsuji, S. and Maeno, Y. and Pfleiderer, C.},
  journal = {Phys. Rev. Lett.},
  volume = {86},
  issue = {12},
  pages = {2661--2664},
  numpages = {0},
  year = {2001},
  month = {Mar},
  publisher = {American Physical Society},
  doi = {10.1103/PhysRevLett.86.2661},
  url = {https://link.aps.org/doi/10.1103/PhysRevLett.86.2661}
}

@Article{Borzi2007,
  author    = {Borzi, R. A. and Grigera, S. A. and Farrell, J. and Perry, R. S. and Lister, S. J. S. and Lee, S. L. and Tennant, D. A. and Maeno, Y. and Mackenzie, A. P.},
  journal   = {Science},
  title     = {Formation of a Nematic Fluid at High Fields in {Sr$_3$Ru$_2$O$_7$}},
  year      = {2007},
  month     = jan,
  number    = {5809},
  pages     = {214--217},
  volume    = {315},
  comment   = {doi: 10.1126/science.1134796},
  doi       = {10.1126/science.1134796},
  publisher = {American Association for the Advancement of Science},
  url       = {https://doi.org/10.1126/science.1134796},
}

@article{Lester2015,
archivePrefix = {arXiv},
arxivId = {1409.7054},
author = {Lester, C. and Ramos, S. and Perry, R. S. and Croft, T. P. and Bewley, R. I. and Guidi, T. and Manuel, P. and Khalyavin, D. D. and Forgan, E. M. and Hayden, S. M.},
doi = {10.1038/nmat4181},
issn = {14764660},
journal = {Nature Materials},
number = {4},
pages = {373--378},
title = {{Field-tunable spin-density-wave phases in {Sr$_3$Ru$_2$O$_7$}}},
volume = {14},
year = {2015}
}

@Article{Settai1997,
  author    = {Settai, Rikio and Misawa, Akira and Araki, Shingo and Kosaki, Masato and Sugiyama, Kiyohiro and Takeuchi, Tetsuya and Kindo, Koichi and Haga, Yoshinori and Yamamoto, Etsuji and Ōnuki, Yoshichika},
  journal   = {J. Phys. Soc. Jpn.},
  title     = {Single Crystal Growth and Magnetic Properties of {CeRh$_2$Si$_2$}},
  year      = {1997},
  issn      = {0031-9015},
  month     = aug,
  number    = {8},
  pages     = {2260--2263},
  volume    = {66},
  comment   = {doi: 10.1143/JPSJ.66.2260},
  doi       = {10.1143/JPSJ.66.2260},
  publisher = {The Physical Society of Japan},
  url       = {https://doi.org/10.1143/JPSJ.66.2260},
}

@article{Abe1997,
  author    = {Abe, Hideki and Suzuki, Hiroyuki and Kitazawa, Hideaki and Matsumo, Takehiko and Kido, Giyuu},
  journal   = {J. Phys. Soc. Jpn.},
  title     = {Successive Field Induced Magnetic Phase Transitions of Heavy Fermion Compound {CeRh$_2$Si$_2$}},
  year      = {1997},
  issn      = {0031-9015},
  number    = {8},
  pages     = {2525--2526},
  volume    = {66},
  comment   = {doi: 10.1143/JPSJ.66.2525},
  doi       = {10.1143/JPSJ.66.2525},
  publisher = {The Physical Society of Japan},
  url       = {https://doi.org/10.1143/JPSJ.66.2525}
}

@article{Knafo2010,
  title = {High-field metamagnetism in the antiferromagnet {${\text{CeRh}}_{2}{\text{Si}}_{2}$}},
  author = {Knafo, W. and Aoki, D. and Vignolles, D. and Vignolle, B. and Klein, Y. and Jaudet, C. and Villaume, A. and Proust, C. and Flouquet, J.},
  journal = {Phys. Rev. B},
  volume = {81},
  issue = {9},
  pages = {094403},
  numpages = {9},
  year = {2010},
  month = {Mar},
  publisher = {American Physical Society},
  doi = {10.1103/PhysRevB.81.094403},
  url = {https://link.aps.org/doi/10.1103/PhysRevB.81.094403}
}

@article{Haen1987,
  author = {P. Haen and J. Flouquet and F. Lapierre and P. Lajay and G. Remenyi},
  title = {{Metamagnetic-like transition in CeRu$_2$Si$_2$}},
  journal = {J. Low Temp. Phys.},
  volume = {67},
  pages = {391},
  year = {1987},
  url = {http://link.springer.com/10.1007/BF00710351}
}

@article{Scheerer2014a,
author = {Scheerer, G. W. and Knafo, W. and Aoki, D. and Nardone, M. and Zitouni, A. and B{\'{e}}ard, J. and Billette, J and Barata, J and Jaudet, C and Suleiman, M. and Frings, P. and Drigo, L. and Audouard, A. and Matsuda, T. D. and Pourret, A. and Knebel, G. and Flouquet, J.},
doi = {10.1103/PhysRevB.89.165107},
issn = {1098-0121},
journal = {Phys. Rev. B},
month = {apr},
number = {16},
pages = {165107},
title = {{Fermi surface in the hidden-order state of {URu$_2$Si$_2$} under intense pulsed magnetic fields up to 81 T}},
url = {https://link.aps.org/doi/10.1103/PhysRevB.89.165107},
volume = {89},
year = {2014}
}

@Article{Aoki2012_URu2Si2,
  author   = {Aoki ,Dai and Knebel ,Georg and Sheikin ,Ilya and Hassinger ,Elena and Malone ,Liam and D. Matsuda ,Tatsuma and Flouquet ,Jacques},
  journal  = {J. Phys. Soc. Jpn.},
  title    = {High-Field Fermi Surface Properties in the Low-Carrier Heavy-Fermion Compound {${\mathrm{URu}}_{2}{\mathrm{Si}}_{2}$}},
  year     = {2012},
  number   = {7},
  pages    = {074715},
  volume   = {81},
  doi      = {10.1143/JPSJ.81.074715},
   url      = {https://doi.org/10.1143/JPSJ.81.074715},
}

@article{Flouquet2002,
author = {Flouquet, J and Haen, P and Raymond, S and Aoki, D and Knebel, G},
doi = {10.1016/S0921-4526(02)01126-2},
issn = {09214526},
journal = {Physica B: Condensed Matter},
number = {1-4},
pages = {251--261},
title = {{Itinerant metamagnetism of CeRu$_2$Si$_2$: Bringing out the dead. Comparison with the new Sr$_3$Ru$_2$O$_7$ case}},
volume = {319},
year = {2002}
}

@article{HAoki2014,
author = {Aoki, Haruyoshi and Kimura, Noriaki and Terashima, Taichi},
doi = {10.7566/JPSJ.83.072001},
issn = {0031-9015},
journal = {J. Phys. Soc. Jpn.},
month = {jul},
number = {7},
pages = {072001},
title = {{Fermi Surface Properties, Metamagnetic Transition and Quantum Phase Transition of {CeRu$_2$Si$_2$} and Its Alloys Probed by the dHvA Effect}},
url = {https://journals.jps.jp/doi/10.7566/JPSJ.83.072001},
volume = {83},
year = {2014}
}

@article{Boukahil2014,
  title = {Lifshitz transition and metamagnetism: Thermoelectric studies of {${\mathrm{CeRu}}_{2}{\mathrm{Si}}_{2}$}},
  author = {Boukahil, M. and Pourret, A. and Knebel, G. and Aoki, D. and \ifmmode \bar{O}\else \={O}\fi{}nuki, Y. and Flouquet, J.},
  journal = {Phys. Rev. B},
  volume = {90},
  issue = {7},
  pages = {075127},
  numpages = {6},
  year = {2014},
  month = {Aug},
  publisher = {American Physical Society},
  doi = {10.1103/PhysRevB.90.075127},
  url = {https://link.aps.org/doi/10.1103/PhysRevB.90.075127}
}

@article{Pourret2019,
title = {Transport Spectroscopy of the Field Induced Cascade of {L}ifshitz Transitions in {YbRh$_2$Si$_2$}},
author = {Pourret, Alexandre and Sharapov, Sergei G. and Matsuda, Tatsuma D. and Knebel, Georg and Zwicknagl, Gertrud and Varlamov, Andrey A.},
issn = {0031-9015},
journal = {J. Phys. Soc. Jpn.},
number = {10},
pages = {104702},
volume = {88},
year = {2019},
doi = {10.7566/jpsj.88.104702},
URL = {https://doi.org/10.7566/JPSJ.88.104702},}

@Article{Geibel1991,
  author   = {Geibel, C. and Schank, C. and Thies, S. and Kitazawa, H. and Bredl, C. D. and Böhm, A. and Rau, M. and Grauel, A. and Caspary, R. and Helfrich, R. and Ahlheim, U. and Weber, G. and Steglich, F.},
  journal  = {Zeitschrift für Physik B Condensed Matter},
  title    = {{Heavy-fermion superconductivity at $T_c=2$~K in the antiferromagnet UPd$_2$Al$_3$}},
  year     = {1991},
  issn     = {1431-584X},
  number   = {1},
  pages    = {1--2},
  volume   = {84},
  doi      = {10.1007/BF01453750},
  refid    = {Geibel1991},
  url      = {https://doi.org/10.1007/BF01453750},
}

@Article{Krimmel1992,
  author   = {Krimmel, A. and Fischer, P. and Roessli, B. and Maletta, H. and Geibel, C. and Schank, C. and Grauel, A. and Loidl, A. and Steglich, F.},
  journal  = {Zeitschrift für Physik B Condensed Matter},
  title    = {{Neutron diffraction study of the heavy fermion superconductors UM$_2$Al$_3$ (M=Pd, Ni)}},
  year     = {1992},
  issn     = {1431-584X},
  number   = {2},
  pages    = {161--162},
  volume   = {86},
  doi      = {10.1007/BF01313821},
  refid    = {Krimmel1992},
  url      = {https://doi.org/10.1007/BF01313821},
}

@article{Zwicknagl2003,
    title = {{The dual nature of 5f electrons and the origin of heavy fermions in U compounds}},
    author = {Zwicknagl, G. and Fulde, P.},
    journal = {J. Phys.: Condens. Matter},
    number = {28},
    volume = {15},
    pages = {S1911},
    year = {2003},
    doi = {10.1088/0953-8984/15/28/302},
    url = {https://10.1088/0953-8984/15/28/302},
}

@article{Zwicknagl2016,
author = {Zwicknagl, Gertrud},
url = {https://10.1088/0034-4885/79/12/124501},
doi = {10.1088/0034-4885/79/12/124501},
issn = {00344885},
journal = {Reports on Progress in Physics},
number = {12},
pages = {124501},
publisher = {IOP Publishing},
title = {{The utility of band theory in strongly correlated electron systems}},
volume = {79},
year = {2016}
}

@article{Blackburn2006,
  title = {Fermi Surface Topology and the Superconducting Gap Function in {${\mathrm{UPd}}_{2}{\mathrm{Al}}_{3}$}: A Neutron Spin-Echo Study},
  author = {Blackburn, E. and Hiess, A. and Bernhoeft, N. and Rheinst\"adter, M. C. and H\"au\ss{}ler, W. and Lander, G. H.},
  journal = {Phys. Rev. Lett.},
  volume = {97},
  issue = {5},
  pages = {057002},
  numpages = {4},
  year = {2006},
  month = {Aug},
  publisher = {American Physical Society},
  doi = {10.1103/PhysRevLett.97.057002},
  url = {https://link.aps.org/doi/10.1103/PhysRevLett.97.057002}
}

@Article{Fujimori2007,
  author   = {Fujimori, Shin Ichi and Saitoh, Yuji and Okane, Tetsuo and Fujimori, Atsushi and Yamagami, Hiroshi and Haga, Yoshinori and Yamamoto, Etsuji and Onuki, Yoshichika},
  journal  = {Nature Physics},
  title    = {{Itinerant to localized transition of f electrons in the antiferromagnetic superconductor {UPd$_2$Al$_3$}}},
  year     = {2007},
  issn     = {17452481},
  number   = {9},
  pages    = {618--622},
  volume   = {3},
  doi      = {10.1038/nphys651},
  url      = {https://10.1038/nphys651},
}

@Article{deVisser1993,
  author   = {A. {de Visser} and K. Bakker and L.T. Tai and A.A. Menovsky and S.A.M. Mentink and G.J. Nieuwenhuys and J.A. Mydosh},
  journal  = {Physica B: Condensed Matter},
  title    = {High-field magnetoresistance of heavy-fermion {UPd$_2$Al$_3$}},
  year     = {1993},
  issn     = {0921-4526},
  pages    = {291-293},
  volume   = {186-188},
  doi      = {https://doi.org/10.1016/0921-4526(93)90557-M},
  url      = {https://www.sciencedirect.com/science/article/pii/092145269390557M},
}

@Article{Oda1994,
  author   = {Oda ,Kiwamu and Kumada ,Takayuki and Sugiyama ,Kiyohiro and Sato ,Noriaki and Komatubara ,Takemi and Date ,Muneyuki},
  journal  = {J. Phys. Soc. Jpn.},
  title    = {High Field Magnetization of {UPd$_2$Al$_3$}},
  year     = {1994},
  number   = {8},
  pages    = {3115-3121},
  volume   = {63},
  doi      = {10.1143/JPSJ.63.3115},
  url      = {https://doi.org/10.1143/JPSJ.63.3115},
}

@Article{deVisser1994,
  author   = {A. {de Visser} and H.P. {van der Meulen} and L.T. Tai and A.A. Menovsky},
  journal  = {Physica B: Condensed Matter},
  title    = {Anisotropy of the antiferromagnetic phase diagram of heavy-fermion {UPd$_2$Al$_3$}},
  year     = {1994},
  issn     = {0921-4526},
  pages    = {100-102},
  volume   = {199-200},
  doi      = {https://doi.org/10.1016/0921-4526(94)91748-5},
  url      = {https://www.sciencedirect.com/science/article/pii/0921452694917485},
}

@Article{Sakon2002,
  author               = {Sakon, T and Sato, NK and Nakanishi, Y and Komatsubara, T and Motokawa, M},
journal              = {Jpn. J. Appl. Phys.},
  title                = {Magnetic phase diagram of {UPd$_2$Al$_3$} in high magnetic fields},
  year                 = {2002},
  issn                 = {0021-4922},
  month                = {JUN},
  number               = {6A},
  pages                = {3673-3677},
  volume               = {41},
  doi                  = {10.1143/JJAP.41.3673},
  orcid-numbers        = {Sakon, Takuo/0000-0002-3343-7108},
  researcherid-numbers = {Komatsubara, Tetsuta/GSI-7012-2022},
  unique-id            = {WOS:000177169500012},
}

@article{Oda1999,
Author = {Oda, K and Sugiyama, K and Sato, NK and Komatsubara, T and Kindo, K and Onuki, Y},
Title = {Metamagnetic transition and phase diagram of a heavy fermion
   superconductor {UPd$_2$Al$_3$}},
Journal = {J. Phys. Soc. Jpn.},
Year = {1999},
Volume = {68},
Number = {9},
Pages = {3115-3116},
Month = {SEP},
DOI = {10.1143/JPSJ.68.3115},
ISSN = {0031-9015},
ResearcherID-Numbers = {Komatsubara, Tetsuta/GSI-7012-2022},
Unique-ID = {WOS:000082909900040},
}

@article{Inada1994,
Author = {Inada, Y and Aono, H and Ishiguru, A and Kimura, J and Sato, N and
   Sawada, A and Komatsubara, T},
Title = {{dHvA Effect on {UPd$_2$Al$_3$}}},
Journal = {Physica B},
Year = {1994},
Volume = {199},
Pages = {119-121},
Month = {APR},
DOI = {10.1016/0921-4526(94)91754-X},
ISSN = {0921-4526},
ResearcherID-Numbers = {Komatsubara, Tetsuta/GSI-7012-2022},
Unique-ID = {WOS:A1994NV53800039},
url = {https//doi.org/10.1016/0921-4526(94)91754-X},
}

@article{Haga1999,
  author = {Y. Haga and Y. Inada and K. Sakurai and Y. Tokiwa and E. Yamamoto and T. Honma and Y. \=Onuki},
  title = {De Haas-van Alphen oscillation in both the normal and superconducting
   mixed states of {UPd$_2$Al$_3$}},
  journal = {J. Phys. Soc. Jpn.},
  volume = {68},
  pages = {342},
  year = {1999},
  DOI = {10.1143/JPSJ.68.342},
  url = {https://10.1143/JPSJ.68.342},
}

@article{Haga2000,
Author = {Haga, Y and Inada, Y and Sakurai, K and Tokiwa, Y and Yamamoto, E and
   Honma, T and \=Onuki, Y},
Title = {{de Haas-van Alphen effect in a heavy fermion superconductor
   {UPd$_2$Al$_3$}}},
Journal = {Physica B},
Year = {2000},
Volume = {284},
Number = {2},
Pages = {1291-1292},
Month = {JUL},
DOI = {10.1016/S0921-4526(99)02566-1},
ISSN = {0921-4526},
ResearcherID-Numbers = {Haga, Yoshinori/C-7679-2011
   Tokiwa, Yoshifumi/P-6593-2015},
ORCID-Numbers = {Haga, Yoshinori/0000-0002-4605-5117
   },
Unique-ID = {WOS:000087423100085},
}

@article{Sandratskii1994,
  title = {{Electronic structure, magnetic, and Fermi-surface properties of ${\mathrm{UPd}}_{2}$${\mathrm{Al}}_{3}$}},
  author = {Sandratskii, L. M. and K\"ubler, J. and Zahn, P. and Mertig, I.},
  journal = {Phys. Rev. B},
  volume = {50},
  issue = {21},
  pages = {15834--15842},
  numpages = {0},
  year = {1994},
  month = {Dec},
  publisher = {American Physical Society},
  doi = {10.1103/PhysRevB.50.15834},
  url = {https://link.aps.org/doi/10.1103/PhysRevB.50.15834}
}

@article{Terashima1997,
author = {Terashima, T and Haworth, C and Takashita, M and Aoki, H},
journal = {Phys. Rev. B},
number = {20},
pages = {369--372},
title = {{Heavy fermions survive the metamagnetic transition in {UPd$_{2}$Al$_{3}$}}},
url = {http://link.aps.org/doi/10.1103/PhysRevB.55.R13369},
volume = {55},
year = {1997}
}

@article{Sugiyama2000,
title = {{Metamagnetism of uranium heavy-fermion compounds UPd$_2$Al$_3$, URu$_2$Si$_2$ and UPt$_3$}},
journal = {Physica B: Condensed Matter},
volume = {281-282},
pages = {244-246},
year = {2000},
issn = {0921-4526},
doi = {https://doi.org/10.1016/S0921-4526(99)00926-6},
url = {https://www.sciencedirect.com/science/article/pii/S0921452699009266},
author = {K Sugiyama and M Nakashima and M Futoh and H Ohkuni and T Inoue and K Kindo and N Kimura and E Yamamoto and Y Haga and T Honma and R Settai and Y Ōnuki}
}

@article{Paolasini1994,
  title = {Field dependence of magnetic structure of {${\mathrm{UPd}}_{2}{\mathrm{Al}}_{3}$} in the normal state},
  author = {Paolasini, L. and Paix\~ao, J. A. and Lander, G. H. and Burlet, P. and Sato, N. and Komatsubara, T.},
  journal = {Phys. Rev. B},
  volume = {49},
  issue = {10},
  pages = {7072(R)--7075(R)},
  year = {1994},
  month = {Mar},
  publisher = {American Physical Society},
  doi = {10.1103/PhysRevB.49.7072},
  url = {https://link.aps.org/doi/10.1103/PhysRevB.49.7072}
}

@article{Knopfle1996,
doi = {10.1088/0953-8984/8/7/014},
url = {https://doi.org/10.1088/0953-8984/8/7/014},
year = {1996},
month = {feb},
publisher = {},
volume = {8},
number = {7},
pages = {901},
author = {K Knöpfle and A Mavromaras and L M Sandratskii and J Kübler},
title = {{The Fermi surface of UPd$_2$Al$_3$}},
journal = {J. Phys.: Condens. Matter}
}

@Article{PalacioMorales2016,
  author        = {{Palacio Morales}, A. and Pourret, A. and Knebel, G. and Bastien, G. and Taufour, V. and Aoki, D. and Yamagami, H. and Flouquet, J.},
  journal       = {Phys. Rev. B},
  title         = {{Thermoelectric power quantum oscillations in the ferromagnet UGe$_2$}},
  year          = {2016},
  issn          = {2469-9969},
  month         = apr,
  number        = {0},
  pages         = {155120},
  volume        = {93},
  doi           = {10.1103/physrevb.93.155120},
   url = {https://link.aps.org/doi/10.1103/PhysRevB.93.155120},
   publisher     = {American Physical Society (APS)},
}

@Article{Pantsulaya1989,
  author   = {A.V. Pantsulaya and A.A. Varlamov},
  journal  = {Physics Letters A},
  title    = {Possibility of observation of giant oscillations of thermoelectric power in normal metal},
  year     = {1989},
  issn     = {0375-9601},
  number   = {6},
  pages    = {317-320},
  volume   = {136},
  doi      = {https://doi.org/10.1016/0375-9601(89)90824-4},
  url      = {https://www.sciencedirect.com/science/article/pii/0375960189908244},
}

@article{Matsuda2000,
author = {Matsuda, TD and Sugawara, H and Aoki, Y and Sato, H},
journal = {Phys. Rev. B},
number = {21},
pages = {852--855},
title = {{Transport properties of the anisotropic itinerant-electron metamagnet UCoAl}},
url = {http://prb.aps.org/abstract/PRB/v62/i21/p13852_1},
volume = {62},
year = {2000}
}

@book{Allen92,
  author = {J. W. Allen},
  title = {Resonant Photoemission of Solids with Strongly Correlated Electrons},
  series = {Synchrotron Radiation Research: Advances in Surface and Interface Science},
  chapter = {6},
  pages = {253},
  publisher = {Plenum Press},
  address = {New York},
  year = {1992},
  volume = {1}
}

@Article{Fujimori99,
  author   = {Fujimori, Shin-ichi and Yasuharu Saito and Masaharu Seki and Koji Tamura and Munenori Mizuta and Ken-ichiro Yamaki and Ken Sato and Tetsuo Okane and Akinori Tanaka and Noriaki Sato and Takemi Komatsubara and Yasuhisa Tezuka and Shik Shin and Shoji Suzuki and Shigeru Sato},
  journal  = {J. Electron Spectrosc. Relat. Phenom.},
  title    = {The {U} {5$f$} states in the heavy fermion uranium compound {UPd$_2$Al$_3$}, studied by resonant and {X}-ray photoelectron spectroscopy},
  year     = {1999},
  issn     = {0368-2048},
  pages    = {439-442},
  volume   = {101-103},
  doi      = {https://doi.org/10.1016/S0368-2048(98)00506-4},
  url      = {https://www.sciencedirect.com/science/article/pii/S0368204898005064},
}

@article{Bohm92,
author = {B\"{o}hm, A. and Grauel, A. and Sato, N. and Schank, C. and Geibel, C. and Komatsubara, T. and Weber, G. and Steglich, F.},
title = {CRYSTALLINE ELECTRIC FIELD EFFECTS IN {UPd$_2$Al$_3$}},
journal = {International Journal of Modern Physics B},
volume = {07},
number = {01n03},
pages = {34-37},
year = {1993},
doi = {10.1142/S0217979293000093},
URL = {https://doi.org/10.1142/S0217979293000093}
}

@article{Grauel92,
  title = {Tetravalency and magnetic phase diagram in the heavy-fermion superconductor {${\mathrm{UPd}}_{2}$${\mathrm{Al}}_{3}$}},
  author = {Grauel, A. and B\"ohm, A. and Fischer, H. and Geibel, C. and K\"ohler, R. and Modler, R. and Schank, C. and Steglich, F. and Weber, G. and Komatsubara, T. and Sato, N.},
  journal = {Phys. Rev. B},
  volume = {46},
  issue = {9},
  pages = {5818(R)--5821(R)},
  numpages = {0},
  year = {1992},
  month = {Sep},
  publisher = {American Physical Society},
  doi = {10.1103/PhysRevB.46.5818},
  url = {https://link.aps.org/doi/10.1103/PhysRevB.46.5818}
}

@incollection{Thalmeier2005a,
  author = {P. Thalmeier and G. Zwicknagl},
  title = {Unconventional superconductivity and magnetism in lanthanide and actinide intermetallic compounds},
  booktitle = {Handbook on the Physics and Chemistry of Rare Earth},
  volume = {34},
  pages = {135--287},
  publisher = {Elsevier B. V.},
  year = {2005}
}

@incollection{Thalmeier2005b,
  author = {P. Thalmeier and G. Zwicknagl and O. Stockert and G. Sparn and F. Steglich},
  title = {},
  booktitle = {Frontiers in Superconducting Materials},
  pages = {109--182},
  publisher = {Springer},
  year = {2005}
}

@incollection{Fulde2006,
  author = {P. Fulde and P. Thalmeier and G. Zwicknagl},
  title = {Strongly correlated electrons},
  booktitle = {Solid State Physics},
  volume = {60},
  pages = {1},
  publisher = {Academic Press},
  year = {2006}
}

@Article{Petit2003,
  author               = {Petit, L and Svane, A and Temmerman, WM and Szotek, Z and Tyer, R},
  journal              = {Europhysics Letters},
  title                = {Ab initio determination of the localized/delocalized $f$-manifold in {UPd$_2$Al$_3$}},
  year                 = {2003},
  issn                 = {0295-5075},
  month                = {MAY},
  number               = {3},
  pages                = {391-397},
  volume               = {62},
  doi                  = {10.1209/epl/i2003-00409-9},
  eissn                = {1286-4854},
  orcid-numbers        = {Petit, Leon/0000-0001-6489-9922 },
  researcherid-numbers = {Petit, Leon/B-5255-2008 Szotek, Zdzislawa/KHW-0164-2024},
  unique-id            = {WOS:000182296200015},
}

@Article{Wills2004,
  author               = {Wills, JM and Eriksson, O and Delin, A and Andersson, PH and Joyce, JJ and Durakiewicz, T and Butterfield, MT and Arko, AJ and Moore, DP and Morales, LA},
  journal              = {J. Electron Spectrosc. Relat. Phenom.},
  title                = {A novel electronic configuration of the 5$f$ states in $\delta$-plutonium as revealed by the photo-electron spectra},
  year                 = {2004},
  issn                 = {0368-2048},
  month                = {APR},
  number               = {2-3},
  pages                = {163-166},
  volume               = {135},
  doi                  = {10.1016/j.elspec.2004.02.169},
  eissn                = {1873-2526},
  orcid-numbers        = {Durakiewicz, Tomasz/0000-0002-1980-1874 Eriksson, Olle/0000-0001-5111-1374},
  researcherid-numbers = {Delin, Anna/P-2100-2014 Eriksson, Olle/E-3265-2014 /K-6432-2013 /L-9769-2019},
  unique-id            = {WOS:000222878900011},
}

@Article{Zwicknagl2003a,
  author   = {Zwicknagl, G. and Yaresko, A. and Fulde, P.},
  journal  = {Phys. Rev. B},
  title    = {{Fermi surface and heavy masses for UPd$_2$Al$_3$}},
  year     = {2003},
  issn     = {1550235X},
  issue    = {5},
  pages    = {052508},
  volume   = {68},
  doi      = {10.1103/PhysRevB.68.052508},
}

@article{Fujimori2012,
author = {Fujimori ,Shin-ichi and Ohkochi ,Takuo and Kawasaki ,Ikuto and Yasui ,Akira and Takeda ,Yukiharu and Okane ,Tetsuo and Saitoh ,Yuji and Fujimori ,Atsushi and Yamagami ,Hiroshi and Haga ,Yoshinori and Yamamoto ,Etsuji and Tokiwa ,Yoshifumi and Ikeda ,Shugo and Sugai ,Takashi and Ohkuni ,Hitoshi and Kimura ,Noriaki and Ōnuki ,Yoshichika},
title = {Electronic Structure of Heavy Fermion Uranium Compounds Studied by Core-Level Photoelectron Spectroscopy},
journal = {J. Phys. Soc. Jpn.},
volume = {81},
number = {1},
pages = {014703},
year = {2012},
doi = {10.1143/JPSJ.81.014703},
URL = {https://doi.org/10.1143/JPSJ.81.014703}
}

@InCollection{Fujimori2014,
  author    = {Fujimori, Shin-ichi and Kawasaki, Ikuto and Yasui, Akira and Takeda, Yukiharu and Okane, Tetsuo and Saitoh, Yuji and Fujimori, Atsushi and Yamagami, Hiroshi and Haga, Yoshinori and Yamamoto, Etsuji and \=Onuki, Yoshichika},
  publisher = {J. Phys. Soc. Jpn.},
  title     = {Angle Resolved Photoelectron Spectroscopy Study of Heavy Fermion Superconductor {UPd$_2$Al$_3$}},
  year      = {2014},
  month     = jun,
  number    = {0},
  series    = {JPS Conference Proceedings},
  volume    = {3},
  comment   = {doi:10.7566/JPSCP.3.011072},
  doi       = {10.7566/JPSCP.3.011072},
  journal   = {Proceedings of the International Conference on Strongly Correlated Electron Systems (SCES2013)},
  url       = {https://doi.org/10.7566/JPSCP.3.011072},
}

@article{Fujimori2016,
doi = {10.1088/0953-8984/28/15/153002},
url = {https://doi.org/10.1088/0953-8984/28/15/153002},
year = {2016},
month = {mar},
publisher = {IOP Publishing},
volume = {28},
number = {15},
pages = {153002},
author = {Fujimori, Shin-ichi},
title = {Band structures of 4f and 5f materials studied by angle-resolved photoelectron spectroscopy},
journal = {J. Phys.: Condens. Matter}
}

@article{Fujimori2019,
Author = {Fujimori, Shin-ichi and Kobata, Masaharu and Takeda, Yukiharu and Okane,
   Tetsuo and Saitoh, Yuji and Fujimori, Atsushi and Yamagami, Hiroshi and
   Haga, Yoshinori and Yamamoto, Etsuji and Onuki, Yoshichika},
Title = {Manifestation of electron correlation effect in $5f$ states of uranium compounds revealed by $4d-5f$ resonant photoelectron spectroscopy},
Journal = {Phys. Rev. B},
Year = {2019},
Volume = {99},
pages = {035109},
Number = {3},
Month = {JAN 4},
DOI = {10.1103/PhysRevB.99.035109},
url = {https://link.aps.org/doi/10.1103/PhysRevB.99.035109},
Article-Number = {035109},
ISSN = {2469-9950},
EISSN = {2469-9969},
ResearcherID-Numbers = {Fujimori, Shinichi/AAS-8071-2020
   Fujimori, Atsushi/G-5099-2014
   Yamagami, Hiroshi/ABA-9022-2021
   Haga, Yoshinori/C-7679-2011},
ORCID-Numbers = {Fujimori, Shinichi/0000-0001-5469-7972
   Qu, Kelsie/0009-0007-1214-9936
   },
Unique-ID = {WOS:000455054800001},
}

@article{Niu2020,
  author = {Q. Niu and G. Knebel and D. Braithwaite and D. Aoki and G. Lapertot and G. Seyfarth and J.-P. Brison and J. Flouquet and A. Pourret},
  title = {Fermi-Surface Instability in the Heavy-Fermion Superconductor {UTe$_{2}$}},
  journal = {Phys. Rev. Lett.},
  volume = {124},
  number = {8},
  pages = {086601},
  month = {Feb},
  year = {2020},
  doi = {10.1103/PhysRevLett.124.086601}
}

@article{Knebel2024,
  title = {$c$-axis electrical transport at the metamagnetic transition in the heavy-fermion superconductor {${\mathrm{UTe}}_{2}$} under pressure},
  author = {Knebel, G. and Pourret, A. and Rousseau, S. and Marquardt, N. and Braithwaite, D. and Honda, F. and Aoki, D. and Lapertot, G. and Knafo, W. and Seyfarth, G. and Brison, J-P. and Flouquet, J.},
  journal = {Phys. Rev. B},
  volume = {109},
  issue = {15},
  pages = {155103},
  numpages = {19},
  year = {2024},
  month = {Apr},
  publisher = {American Physical Society},
  doi = {10.1103/PhysRevB.109.155103},
  url = {https://link.aps.org/doi/10.1103/PhysRevB.109.155103}
}

@article{Freericks1992,
  title = {Heavy-fermion systems in magnetic fields: The metamagnetic transition},
  author = {Freericks, J. K. and Falicov, L. M.},
  journal = {Phys. Rev. B},
  volume = {46},
  issue = {2},
  pages = {874--879},
  numpages = {0},
  year = {1992},
  month = {Jul},
  publisher = {American Physical Society},
  doi = {10.1103/PhysRevB.46.874},
  url = {https://link.aps.org/doi/10.1103/PhysRevB.46.874}
}

@article{Inada1999,
author = {Inada ,Yoshihiko and Yamagami ,Hiroshi and Haga ,Yoshinori and Sakurai ,Kenji and Tokiwa ,Yoshihumi and Honma ,Tetsuo and Yamamoto ,Etsuji and Ōnuki ,Yoshichika and Yanagisawa ,Takashi},
title = {{Fermi Surface and de Haas-van Alphen Oscillation in both the Normal and Superconducting Mixed States of UPd$_2$Al$_3$}},
journal = {J. Phys. Soc. Jpn.},
volume = {68},
number = {11},
pages = {3643-3654},
year = {1999},
doi = {10.1143/JPSJ.68.3643},
URL = {https://doi.org/10.1143/JPSJ.68.3643}
}

@article{Palacio2015,
  title = {Fermi surface instabilities in {${\mathrm{CeRh}}_{2}{\mathrm{Si}}_{2}$} at high magnetic field and pressure},
  author = {{Palacio Morales}, A. and Pourret, A. and Seyfarth, G. and Suzuki, M.-T. and Braithwaite, D. and Knebel, G. and Aoki, D. and Flouquet, J.},
  journal = {Phys. Rev. B},
  volume = {91},
  issue = {24},
  pages = {245129},
  numpages = {9},
  year = {2015},
  month = {Jun},
  publisher = {American Physical Society},
  doi = {10.1103/PhysRevB.91.245129},
  url = {https://link.aps.org/doi/10.1103/PhysRevB.91.245129}
}

@article{Pfau2012,
  title = {Thermoelectric transport across the metamagnetic transition of {CeRu${}_{2}$Si${}_{2}$}},
  author = {Pfau, Heike and Daou, Ramzy and Brando, Manuel and Steglich, Frank},
  journal = {Phys. Rev. B},
  volume = {85},
  issue = {3},
  pages = {035127},
  numpages = {5},
  year = {2012},
  month = {Jan},
  publisher = {American Physical Society},
  doi = {10.1103/PhysRevB.85.035127},
  url = {https://link.aps.org/doi/10.1103/PhysRevB.85.035127}
}

@article{Palacio2013,
  title = {Metamagnetic Transition in {UCoAl} Probed by Thermoelectric Measurements},
  author = {{Palacio Morales}, A. and Pourret, A. and Knebel, G. and Combier, T. and Aoki, D. and Harima, H. and Flouquet, J.},
  journal = {Phys. Rev. Lett.},
  volume = {110},
  issue = {11},
  pages = {116404},
  numpages = {5},
  year = {2013},
  month = {Mar},
  publisher = {American Physical Society},
  doi = {10.1103/PhysRevLett.110.116404},
  url = {https://link.aps.org/doi/10.1103/PhysRevLett.110.116404}
}

@article{Bastien2016,
  title = {Lifshitz Transitions in the Ferromagnetic Superconductor {UCoGe}},
  author = {Bastien, Ga\"el and Gourgout, Adrien and Aoki, Dai and Pourret, Alexandre and Sheikin, Ilya and Seyfarth, Gabriel and Flouquet, Jacques and Knebel, Georg},
  journal = {Phys. Rev. Lett.},
  volume = {117},
  issue = {20},
  pages = {206401},
  numpages = {5},
  year = {2016},
  month = {Nov},
  publisher = {American Physical Society},
  doi = {10.1103/PhysRevLett.117.206401},
  url = {https://link.aps.org/doi/10.1103/PhysRevLett.117.206401}
}

@article{Varlamov1989,
  author = {Varlamov, A.A. and Egorov, V.S. and Pantsulaya, A.V.},
  title = {Kinetic properties of metals near electronic topological transitions
	(2 1/2-order transitions)},
  journal = {Advances in Physics},
  year = {1989},
  volume = {38},
  pages = {469-564},
  number = {5},
  doi = {10.1080/00018738900101132},
  url = {http://www.tandfonline.com/doi/abs/10.1080/00018738900101132}
}

@article{Niu2020b,
  title = {{Evidence of Fermi surface reconstruction at the metamagnetic transition of the strongly correlated superconductor ${\mathrm{UTe}}_{2}$}},
  author = {Niu, Q. and Knebel, G. and Braithwaite, D. and Aoki, D. and Lapertot, G. and Vali\ifmmode \check{s}\else \v{s}\fi{}ka, M. and Seyfarth, G. and Knafo, W. and Helm, T. and Brison, J.-P. and Flouquet, J. and Pourret, A.},
  journal = {Phys. Rev. Res.},
  volume = {2},
  issue = {3},
  pages = {033179},
  numpages = {10},
  year = {2020},
  month = {Aug},
  publisher = {American Physical Society},
  doi = {10.1103/PhysRevResearch.2.033179},
  url = {https://link.aps.org/doi/10.1103/PhysRevResearch.2.033179}
}

@article{Varlamov2021,
    author = {Varlamov, A. A. and Galperin, Y. M. and Sharapov, S. G. and Yerin, Yuriy},
    title = {Concise guide for electronic topological transitions},
    journal = {Low Temperature Physics},
    volume = {47},
    number = {8},
    pages = {672-683},
    year = {2021},
    month = {08},
    issn = {1063-777X},
    doi = {10.1063/10.0005556},
    url = {https://doi.org/10.1063/10.0005556}
}

@article{Christensen84,
    author = {Christensen, N. Egede},
    title = {Relativistic band structure calculations},
    journal = {Int. J. Quantum Chem.},
    volume = {25},
    number = {1},
    pages = {233-261},
    year = {1984},
    doi = {10.1002/qua.560250119},
    url = {https://doi.org/10.1002/qua.560250119},

}

@article{Fulde81,
  title = {{Excitonic Mass Enhancement in Praseodymium}},
  author = {White, R. M. and Fulde, P.},
  journal = {Phys. Rev. Lett.},
  volume = {47},
  issue = {21},
  pages = {1540--1542},
  numpages = {0},
  year = {1981},
  month = {Nov},
  publisher = {American Physical Society},
  doi = {10.1103/PhysRevLett.47.1540},
  url = {https://link.aps.org/doi/10.1103/PhysRevLett.47.1540}
}

@article{Knafo2012,
  title = {High-field moment polarization in the ferromagnetic superconductor {UCoGe}},
  author = {Knafo, W. and Matsuda, T. D. and Aoki, D. and Hardy, F. and Scheerer, G. W. and Ballon, G. and Nardone, M. and Zitouni, A. and Meingast, C. and Flouquet, J.},
  journal = {Phys. Rev. B},
  volume = {86},
  issue = {18},
  pages = {184416},
  numpages = {4},
  year = {2012},
  month = {Nov},
  publisher = {American Physical Society},
  doi = {10.1103/PhysRevB.86.184416},
  url = {https://link.aps.org/doi/10.1103/PhysRevB.86.184416}
}

@article{palacio2013b,
author = {Pourret ,Alexandre and Palacio-Morales ,Alexandra and Kr\"{a}mer ,Steffen and Malone ,Liam and Nardone ,Marc and Aoki ,Dai and Knebel ,Georg and Flouquet ,Jacques},
title = {{Fermi Surface Reconstruction inside the Hidden Order Phase of URu$_2$Si$_2$ Probed by Thermoelectric Measurements}},
journal = {J. Phys. Soc. Jpn.},
volume = {82},
number = {3},
pages = {034706},
year = {2013},
doi = {10.7566/JPSJ.82.034706},
URL = {https://doi.org/10.7566/JPSJ.82.034706}
}

@Article{Leenen2024,
  author    = {Leenen, R. and Aoki, D. and Knebel, G. and Pourret, A. and McCollam, A.},
  journal   = {Phys.~Rev.~Res.},
  title     = {Fermi surface and {Lifshitz} transtions of a ferromagnetic superconductor under external magnetic fields},
  year      = {2024},
  issn      = {2643-1564},
  month     = oct,
  number    = {4},
  pages     = {043024},
  volume    = {6},
  date      = {2024-10-08},
  day       = {8},
  doi       = {10.1103/physrevresearch.6.043024},
  publisher = {American Physical Society (APS)},
}

\end{document}